\documentstyle[preprint,aps,epsfig, tighten]{revtex}  % DON'T CHANGE
\newcommand{\beq}{\begin{equation}}
\newcommand{\eeq}{\end{equation}}
\newcommand{\beqa}{\begin{eqnarray}}
\newcommand{\eeqa}{\end{eqnarray}}
\newcommand{\Psib}{\overline{\Psi}}

\newcommand{\Lcal}{{\cal L}}
\newcommand{\p}{\partial}
\begin{document}                % INITIALIZE - DONT CHANGE
\draft
\title{Density dependent hadron field theory for hypernuclei}
\author{C. M. Keil, F. Hofmann, H. Lenske}
\address{Institut f\"ur Theoretische Physik, Universit\"at Gie\ss en,
         Heinrich-Buff-Ring 16, 35392 Gie\ss en, Germany}
% \author{}   % Use this and the next line only if there is a second
% \address{Another University, etc.}  % address. (Remove the left % marks)
\date{\today}
\maketitle
\begin{abstract}                % DON'T CHANGE THIS LINE

The Density Dependent Relativistic Hadron Field (DDRH) theory,
previously introduced and applied to isospin nuclei, is extended to
hypernuclei by including the octet hyperons. Infinite matter
Dirac-Brueckner theory for octet baryons and the derivation of in-medium
DDRH baryon-meson vertices is discussed. From the properties of
Dirac-Brueckner interactions it is found that hyperon and nucleon
self-energies and vertices are related by the ratios of free space
coupling constants. This leads to simple scaling laws for the
in-medium hyperon and nucleon vertices. The model is applied in
relativistic DDRH mean-field calculations to single $\Lambda$ nuclei.
Free space N$\Lambda$ T-matrix results are used for the  scalar vertex.
As the only free parameter the hyperon vector vertex scaling factor is
adjusted to a selected set of hypernuclear data. Spectroscopic data of
single $\Lambda$ hypernuclei over the full mass range are well
described. The reduced $\Lambda$ spin-orbit splitting is reproduced and
found to be related closely the medium dependence of scalar and vector
interactions.

\end{abstract} % % %

%%%%%%%%%%%%%%%%%%%%%%%%%%%%%%%%%%%%%%%%%%%%%%%%%%%%%%%%%%%%%%%%%%%%%%%%%%%%%%%
%%%%%%%%%%%%%%%%%%%%%%%%%%%%%%%%%%%%%%%%%%%%%%%%%%%%%%%%%%%%%%%%%%%%%%%%%%%%%%%
\section{Introduction}               % Introduction goes below.
\label{sec:Intro}

Hypernuclei are unique in providing access to the dynamics of the full
meson and baryon SU(3) flavour octets. Their study is the natural
extension of the isospin dynamics in non-strange nuclei towards a more
general theory of flavour dynamics in a baryonic environment. Obviously,
from a QCD point of view hypernuclei as also isospin nuclei are deep in
the non-perturbative low energy-momentum regime. Hence, a description in
terms of mesons and baryons should be adequate. Single $\Lambda$
hypernuclei, produced in ($K^-,\pi^-$) or ($\pi^+,K^+$) reactions on a
nuclear neutron are the best studied examples. Their properties confirm
that adding a unit of strangeness to an isospin nucleus indeed produces
a system which, to a large extent, follows similar rules as isospin
nuclei \cite{Ba90,Ch89}. Such observations give strong evidence that the
strangeness content of a hypernucleus is in fact stored in a hyperon.
Moreover, hypernuclear spectroscopy indicates the exsistence of shell
structures compatible with independent (quasi-) particle motion in a
static mean-field. The effective potential, however, is found to be
considerably more shallow than for nucleons. A natural explanation for
the reduction in depth to about 50\% of the nucleon value is provided by
assuming that the mean-field producing $\sigma$ and $\omega$ meson
fields are not coupled to strangeness. Under these "ideal mixing"
conditions the meson-hyperon coupling should evolve according to the
ratio of strange to non-strange quarks in a baryon, i.e. a reduction of
vertices by at least a factor of R$_{\sigma,\omega}\sim$2/3 is expected
for $\Lambda$ and $\Sigma$ hyperons. However, such a naive quark
counting model is unable to account for the experimentally observed
decrease of the spin-orbit splitting in $\Lambda$ nuclei.

Modern approaches to hypernuclear structure are using non-relativistic
and relativistic microscopic descriptions. Relativisitic mean-field
(RMF) theories of Walecka-type \cite{SW86} have been applied
successfully  \cite{Ru90,Gl92,Vr98} with empirically adjusted
meson-hyperon vertices. SU(3)$_f$-symmetric field theories incorparating
chirality \cite{Mu99,Pa99} or accounting for the quark structure of
hadrons \cite{Ts98} have been formulated and applied to hypernuclei.
Extensions to still unobserved multi-strangeness systems ($|S|>$2) have
been explored predicting a gain of binding energy when adding a few
units of strangeness to an isospin nucleus \cite{Ru90,Gr96}. The production of strangelets in ultra-relativistic heavy ion collisions as a new form of hadronic matter has been postulated \cite{CG87}. Using
SU(3)$_f$ arguments nucleon-hyperon and hyperon-hyperon interactions in
free space \cite{Re96,St99a,Ri99} and in a nuclear environment
\cite{St99b,Sc98} were calculated.

In this paper, hypernuclei are described in the Density Dependent
Relativistic Hadron (DDRH) theory which was introduced previously as an
effective field theory for isospin nuclei \cite{LF95,FL95}. In DDRH
theory the medium dependence of nuclear interactions is described by
meson-nucleon vertices which are functionals of the fermion field
operators. Lorentz-invariance, thermodynamical consistency and
covariance of the field equations are retained. Taking the functional
dependence of the vertices on density from infinite matter
Dirac-Brueckner Hartree-Fock (DBHF) calculations a practically
parameter free model Lagrangian is obtained once a free space
nucleon-nucleon (NN) interaction is chosen.

A particular conceptual difference to other approaches is the DDRH
treatment of non-linearities in terms of invariants of fermion
field operators. Since the baryon fields are treated as quantum
fields even in the mean-field limit a well defined class of
quantum fluctuations with non-vanishing ground state expectation value
is taken into account \cite{FL95,Ne82}. Dynamically, they
contribute to the Dirac equations as rearrangement self-energies
describing the static polarization of the medium. In standard
relativistic mean-field (RMF) theory non-linearities are
attributed to higher order self-interactions of meson fields
\cite{Ga90}. In mean-field approximation, mesons are treated as
classical fields and fluctuations around the classical field
configurations are neglected by definition. In bulk quantities, as
for example total binding energies, the differences of the DDRH
and the RMF approach are hardly detectable because the DDRH
rearrangement self-energies are cancelled exactly in extensive
thermodynamical quantities \cite{FL95}. But in single particle
quantities like separation energies, wave functions and density
matrices the differences become visible \cite{FL95}. DDRH coupling
constants in asymmetric matter \cite{Jo98a} and modifications from
vacuum polarization \cite{Jo98b} have been investigated.

The extension of DDRH theory to strange baryons is discussed in
sect.~\ref{sec:DensDepNucHyp}. The theoretical formulation is kept
general allowing to include the lowest SU(3) baryon and meson octetts.
By practical reasons, however, only strangeness-neutral meson fields are
taken into account at present. As the central theoretical result we
derive in sect.~\ref{ssec:BBvert} scaling laws for in-medium hyperon
vertices given by an almost density independent renormalization of the
nucleon vertices through the ratios of free space coupling constants. In
phenomenological RMF approaches \cite{Ru90,Gl92,Ma96}, a similar scaling
{\em ansatz} for the meson-$\Lambda$ vertices is used but here it is
obtained theoretically. Equations of motion and the Hartree mean-field
limit are derived. In sect.~\ref{sec:LamNMod} a reduced model appropriate
for single $\Lambda$ nuclei, is introduced. Since strangeness carrying
mean-fields can be neglected (they are of the order ${\cal O}(1/A)$, A the mass
number) the mean-field equations are considerably simplified. In the
applications the $\sigma$ coupling is taken from a theoretical
N$\Lambda$ T-matrix \cite{Re96,Ha98} while the $\omega$ coupling is
determined empirically. DDRH mean-field results for hypernuclei are
presented in sect.~\ref{sec:RelHarDes} and compared to data and
conventional RMF calculations. Spectroscopic data are well described,
reproducing also the reduced spin-orbit splitting in $\Lambda$ nuclei.
The paper closes with a summary and conclusions in
sect.~\ref{sec:summary}.

%%%%%%%%%%%%%%%%%%%%%%%%%%%%%%%%%%%%%%%%%%%%%%%%%%%%%%%%%%%%%%%%%%%%%%%%%%%%%%%
%%%%%%%%%%%%%%%%%%%%%%%%%%%%%%%%%%%%%%%%%%%%%%%%%%%%%%%%%%%%%%%%%%%%%%%%%%%%%%%
\section{Density dependent hadron field theory with hyperons}
\label{sec:DensDepNucHyp}

%%%%%%%%%%%%%%%%%%%%%%%%%%%%%%%%%%%%%%%%%%%%%%%%%%%%%%%%%%%%%%%%%%%%%%%%%%%%%%%
\subsection{The model Lagrangian}
\label{ssec:ModelLagr}

The derivation of a symmetry-broken physical model from a SU(3)$_f$
Lagrangian has been excercised e.g. in ref.~\cite{ChengLi92}.
However, in order to describe nuclear structure phenomena one finds that
most of the SU(3)$_f$ structures are actually not contributing. The
reason is obvious because parity conservation inhibits the appearance of
condensed pseudoscalar fields in nuclei. Hence, neither of the $0^-$
meson fields contributes directly to a hypernuclear calculation, except
through exchange interactions from antisymmetrization. From the $1^-$
vector meson octett condensed isoscalar $\omega$ and isovector $\rho$
meson fields will evolve. In a system with a large fraction of hyperons
also condensed octett $K^*$ and singlett $\Phi$ mesons fields can
appear. However, an apparent shortcoming of a pure SU(3)$_f$ approach is
the missing of scalar mesons and, hence, the absence of a binding
mean-field. A satisfactory description of the $0^+$ meson channels, e.g.
in terms of dynamical 2-meson correlations \cite{Re96,Ha98}, is an
unsolved question.

Here, we follow the line of relativistic mean-field theory \cite{SW86} and
restrict the model to the degrees of freedom which are relevant for the
nuclear structure problem. In practice, we use the
DDRH Lagrangian \cite{FL95} which is extended in the baryon sector
by including the lowest $\text{S} =-1$ ($\Lambda, \Sigma$) and $\text{S} =-2$
($\Xi$) baryons.
We introduce the flavour spinor $\Psi_F$
\beq
\Psi_F = \left( \Psi_N, \Psi_\Lambda, \Psi_\Sigma, \Psi_\Xi \right)^{\text{T}}
\label{eq:FlSp}
\eeq
being composed of the isospin multiplets
\beq
\Psi_N = \left( \begin{array}{c} \psi_p \\ \psi_n \end{array} \right),\quad
\Psi_\Lambda = \psi_\Lambda,\quad
\Psi_\Sigma = \left( \begin{array}{c}
\psi_{\Sigma^+} \\ \psi_{\Sigma^0} \\ \psi_{\Sigma^-} \end{array} \right),\quad
\Psi_\Xi = \left( \begin{array}{c}
\psi_{\Xi^0} \\ \psi_{\Xi^-} \end{array} \right)
\eeq
where $\psi_i$ are Dirac spinors. The full Lagrangian is structured in an
isospin symmetric way.
In the exchange particle sector the isoscalar $\sigma$, $\sigma_s$ ($\equiv$
scalar
$s\overline{s}$ condensate), $\omega$ and $\phi$ meson, the isovector
$\rho$ meson and the photon $\gamma$ are included.
This leads to the Lagrangian
\beqa\label{Lagrangian}
\Lcal &=& \Lcal_{B} + \Lcal_{M} + \Lcal_{int} \nonumber\\
\Lcal_{B} &=& \Psib_{F} \left[ i\gamma_\mu\p^\mu
                              - \hat{M} \right] \Psi_{F} \nonumber\\
\Lcal_{M} &=&\frac{1}{2} \sum_{i = \sigma, \sigma_s}
\left(\p_\mu\Phi_i\p^\mu\Phi_i - m_{\Phi_i}^2\Phi_i^2\right)
           - \frac{1}{2} \sum_{\kappa = \omega, \phi, \rho, \gamma}
             \left( \frac{1}{2} F^{(\kappa)^2} - m_\kappa^2 A^{(\kappa)^2}
\right) \label{eq:ModLagr} \\
\Lcal_{int} &=& \Psib_F \hat{\Gamma}_\sigma(\Psib_F, \Psi_F) \Psi_F \sigma
- \Psib_F \hat{\Gamma}_\omega(\Psib_F, \Psi_F) \gamma_\mu \Psi_F \omega^\mu
- \frac{1}{2}\Psib_F \hat{\vec\Gamma}_\rho(\Psib_F, \Psi_F) \gamma_\mu \Psi_F
\vec\rho^\mu \nonumber\\
&&+ \Psib_F \hat{\Gamma}_{\sigma_s}(\Psib_F, \Psi_F) \Psi_F \sigma_s
- \Psib_F \hat{\Gamma}_\phi(\Psib_F, \Psi_F) \gamma_\mu \Psi_F \phi^\mu
- e \Psib_F \hat{Q} \gamma_\mu \Psi_F A^\mu, \nonumber
\eeqa
where $\Lcal_B$ and $\Lcal_M$ are the free baryonic and  mesonic
Lagrangians, respectively,
and the diagonal matrix $\hat{M}$ contains the free-space baryon
masses.
The meson-baryon interactions are contained in $\Lcal_{int}$.
\beq
F^{(\kappa)}_{\mu\nu} = \p_\mu A_\nu^{(\kappa)} - \p_\nu A_\mu^{(\kappa)}
\label{eq:Fmunu}
\eeq
is the field strength tensor of either the vector mesons ($\kappa = \omega,
\phi, \rho$) or
the photon ($\kappa = \gamma$). In eq.~(\ref{eq:ModLagr}) contractions of the
field
strength tensors are abbreviated as $F^2 = F_{\mu\nu} F^{\mu\nu}$ etc..
$\hat{Q}$ is the electric charge operator.
$\sigma_s$ and $\Phi$ meson fields are included mainly for reasons of
completeness.
The corresponding classical, condensed fields will be important only in
hypermatter
with a significant strangness content, i.e. for hyperon and nucleon fractions of
comparable
order. For single $\Lambda$ nuclei they should be negligible.
Lagrangians of a similar structure, but
with constant meson-baryon vertices, are used successfully in
relativistic mean-field calculations of hypernuclei, see e.g. \cite{Ru90,Gl92}.

An isospin symmetric interaction is obtained with vertices chosen as:
\beq
\label{eq:GammaDef}
\begin{array}{l}
\left( \hat{\Gamma}_\alpha \right)_{BB'} =
\Gamma_{\alpha B}\delta_{BB'},
\quad \alpha = \sigma, \sigma_s, \omega, \phi \\
\left( \hat{\vec\Gamma}_\rho \right)_{BB'} =
\vec{\Gamma}_{\rho B} \delta_{BB'},
\quad \vec{\Gamma}_{\rho B} = \Gamma_{\rho B} \vec{\tau}^B
\end{array}
\qquad \text{for } B, B' = N, \Lambda, \Sigma, \Xi,
\eeq
where $\vec\tau^B$ are the isospin Pauli-matrices.
In DDRH theory the vertices $\Gamma_{\alpha B}=\Gamma_{\alpha B}(\Psib,\Psi)$
are
taken as functionals of the baryon field operators \cite{LF95,FL95}.

%%%%%%%%%%%%%%%%%%%%%%%%%%%%%%%%%%%%%%%%%%%%%%%%%%%%%%%%%%%%%%%%%%%%%%%%%%%%%%%
\subsection{Baryon-baryon vertices from Dirac-Brueckner theory}
\label{ssec:BBvert}

In order to understand the subtleties of including baryon-baryon (BB)
correlations into a field theory of a higher flavour content we briefly
sketch the derivation of the vertices $\Gamma$ from Dirac-Brueckner
theory. The main outcome of the discussions is that nucleon and
hyperon dynamics should be related to a good approximation by
simple scaling laws.

A Lagrangian of the type as defined above leads to a ladder kernel
V$^{BB'}$(q',q) given in momentum represention by the superposition of
one boson exchange (OBE) potentials V$^{BB'}_\alpha$(q',q). The latter
are Lorentz-invariants
\beq\label{obe}
V^{BB'}_\alpha(q',q)=g_{\alpha B} g_{\alpha B'} D_\alpha(t)\langle
\bar{u}^{B_1}(q')
\kappa_\alpha u^{B_3}(q)\rangle \cdot
\langle \bar{u}^{B'_2}(-q') \kappa^\alpha u^{B'_4}(-q)\rangle
\eeq
where t=$(q'-q)^2$ is the 4-momentum transfer, $\kappa_\alpha$ denotes
the Dirac and flavour structure of the vertex with bare coupling
constants $g_{\alpha B}$ and $g_{\alpha B'}$ for baryons $B$ and $B'$
belonging to different isospin multiplets. The Dirac spinors are
indicated by $u^B(q)$. Contractions over Dirac and flavour indices are
indicated by the notation $\kappa_\alpha \cdot \kappa^\alpha$. Working
in the $BB'$ center-of-momentum frame q and q' denote the relative
4-momenta in the in- and outgoing channels $(B_3,$ $B'_4)$ and $(B_1,$
$B'_2)$, respectively, where -q=(q$_0$,-{\bf q}).

Solving the Bethe-Salpeter equation with the kernel
V$^{BB'}=\sum_\alpha{V^{BB'}_\alpha}$
\beq\label{BBS}
{\cal R}^{BB'}(q',q)=V^{BB'}(q',q) +
\int{dk~V^{BB'}(q',k)GQ_F(BB')(k,q){\cal R}^{BB'}(k,q)}\quad .
\eeq
leads to the in-medium interactions ${\cal R}^{BB'}=\langle
k^B_1(q')k^{B'}_2(-q')|{\cal R}(k^B_F,k^{B'}_F)|k^B_3(q)k^{B'}_4(-q) \rangle$.
The Pauli-projected intermediate two-particle propagation, denoted by
$GQ_F(BB')$, introduces an intrinsic dependence on the Fermi momenta
(k$^B_F$,k$^{B'}_F$). Dependencies on the (conserved) total
center-of-mass energy $s=(k^B_3+k^{B'}_4)^2$ are implicit. Evaluating
eq.~(\ref{BBS}) with self-consistent in-medium spinors, including the
self-energies $\Sigma^B(k)$, introduces additional medium dependencies.

In structure calculations the G-matrices ${\cal R}^{BB'}$ are required
in the nuclear matter rest frame rather than in the 2-body c.m. system.
In practice, the transformation is achieved by projection on the
standard set of scalar (S), vector (V), tensor (T), axial vector (A) and
pseudo scalar (P) Lorentz invariants \cite{HS83,HS87,HM87}. For our
purpose, however, a more convenient representation is obtained by
forming appropriate linear combinations of the so obtained coefficients
and to map the set (S,V,T,A,P) onto the vertices of OBE interactions,
eq.~(\ref{obe}). This allows to express the ($BB'$) G-matrices,
eq.~(\ref{BBS}), in terms of renormalized OBE interactions,
\beq\label{robe}
{\cal R}^{BB'}_\alpha(q',q)=z^{BB'}_\alpha(s,t,u|k^B_Fk^{B'}_F) g_{\alpha B}
g_{\alpha B'}
D_\alpha(t)\langle \bar{u}^{B_1}(q') \kappa_\alpha u^{B_3}(q)\rangle \cdot
\langle \bar{u}^{B'_2}(-q') \kappa^\alpha u^{B'_4}(-q)\rangle
\eeq
The vertex invariants have been decomposed further into boson
propagators $D_\alpha(t)$ and renormalization coefficients $z^{BB'}$
which both are Lorentz invariants. From this representation it is
apparent that correlations are shifted into the vertex factors
$z^{BB'}$. In principle, they may depend on the full set of Mandelstam
variables s, t and u and the Fermi momenta of baryons $B$ and $B'$.
However, most of the t dependence is already accounted for by the meson
propagator $D_\alpha(t)$ and considering the mild variation of DB
G-matrices on the center-of-mass energy the z-coefficients can be
expected to depend mainly on the Fermi momenta. If necessary, the u
dependence can be removed to a large extent by adding a term
proportional to $D_\alpha(u)$, i.e. introducing antisymmetrization
explicitly. Note, that antisymmetrization effects contribute only to
states within the same isospin multiplet, i.e. $B=B'$. The self-energy
of baryon $B$, however, includes contributions from all multiplets:
\beqa\label{DBself}
\langle \bar{u}^B(k) && \kappa^{\alpha}
u^B(k)\rangle \Sigma^B_\alpha(k|k_F)= \nonumber\\
&&\sum_{B'}{\int_{K^{B'}_F}{d^4q\left( Tr(\langle k^B q^{B'}|{\cal R}^{BB'}|k^B
q^{B'}\rangle
S^{B'}_F(q))
+ \delta_{BB'}
\langle k^B q^{B'}|{\cal R}^{BB'}|q^B k^{B'}\rangle S^{B'}_F(q) \right)}}
\eeqa
where the space-like integration extends over the Fermi spheres
$K^{B'}_F$ of baryons B' with in-medium (positive energy) propagators
$S^{B'}_F(q)$ and k$_F=(k^N_F,k^Y_F)$ denotes the set of nucleon and
hyperon Fermi momenta.

A particularly appealing aspect of eq.~(\ref{robe}) is that the
z-coefficients can be considered as medium-dependent renormalization
factors of the OBE vertices. Exploiting the fact that the SU(3)$_f$
isospin multiplets are not mixed by strong interactions we are allowed
to assume separability
\beq\label{separ}
z^{BB'}_\alpha(k,q|k^B_F,k^{B'}_F)\simeq
s^B_\alpha(k^B_F)s^{B'}_\alpha(k^{B'}_F)
\eeq
and to neglect the (weak) residual momentum dependence. As a consequence,
\beq\label{rvtx}
g_{\alpha B} z^{BB'}_\alpha(k,q)g_{\alpha B'} \simeq \Gamma_{\alpha B}(k^B_F)
\Gamma_{\alpha B'}(k^{B'}_F) \quad ,
\eeq
where
\beq\label{GammaB}
\Gamma_{\alpha B}(k^B_F) \equiv g_{\alpha B} s^B_\alpha(k^B_F)
\eeq
defines the renormalized in-medium vertices in ladder approximation.
We introduce the antisymmetrized condensed Dirac Hartree-Fock (DHF) meson fields
\beqa\label{DBfields}
\phi_\alpha(k|k_F,\Gamma)&=&
\sum_{B'}{\Gamma_{\alpha B'}D_\alpha(0)
\int_{K^{B'}_F}{dq~Tr(\langle \bar{u}^{B'}(q) \kappa_\alpha
u^{B'}(q)\rangle)}}\\  \nonumber
&+&\sum_{\mu}{\int_{K^{B}_F}{dq~f_{\alpha \mu}\Gamma_{\mu B}D_{\mu}(k-q)
\langle \bar{u}^{B}(q) \kappa_\alpha u^{B}(q)\rangle}}
\eeqa
where $f_{m\mu}$ denotes the Fierz matrix \cite{HS87,BD65} and
$\Gamma=(\Gamma^N,\Gamma^Y)$.
With our choice of momentum independent, global vertices,
eq.~(\ref{DBself}) takes then the approximate form
\beq\label{Gphi}
\Sigma^B_\alpha(k|k^N_F,k^Y_F) \simeq \Gamma_{\alpha B}(k^B_F)
\phi_\alpha(k|k^N_F,k^Y_F,\Gamma)
\eeq
This equation establishes the link to the DDRH Lagrangian:
In DHF approximation
self-energies of the same structure are obtained from eq.~(\ref{Lagrangian}).

In order to derive a self-contained model
we apparently have to introduce a "renormalization" scheme.
This is achieved by choosing symmetric hyper matter, i.e. $k^N_F=k^Y_F=k_F$.
Writing down for that case eq.~(\ref{Gphi}) for nucleons and hyperons explicitly
one finds
\beq\label{YNvertex}
\Gamma_{\alpha Y}(k^Y_F)=\Gamma_{\alpha N}(k^Y_F)\frac{\Sigma^Y_\alpha(k|k_F)}
{\Sigma^N_\alpha(k|k_F)}\vert_{k=k_F,k^N_F=k^Y_F}.
\eeq
which is exact in Hartree approximation.
The above relation is the central result of this section.  The obvious medium
dependencies introduced in eqs.~(\ref{DBself}) and (\ref{Gphi}) by the
external baryon lines are eliminated such that the intrinsic medium
properties of the underlying interactions are projected out. This is
seen more clearly by considering the diagrammatic structure of the DB
self-energies
\cite{HS83,HS87}. The leading order (Hartree) contribution is given by tadpole
diagrams and from a perturbation series expansion in the bare coupling
constants $g_{\alpha B}$
\beq\label{RY}
R^Y_\alpha=\frac{\Sigma^Y_\alpha}{\Sigma^N_\alpha}\simeq
\frac{g_{\alpha Y}}{g_{\alpha N}}
(1 +{\cal O}(1-\frac{M_N}{M_Y}))+\cdots
\eeq
where the realistic case $g_{\alpha Y} < g_{\alpha N}$ is considered. Hence, the
$R^Y_\alpha$ are expected to be state-independent, universal constants whose
values are close to the ratios of the bare coupling constants. For
asymmetric matter with a hyperon fraction
$\zeta_Y=\frac{\rho^Y}{\rho^N} \ll 1$ a corresponding diagrammatic
analysis shows that asymmetry terms are in fact suppressed because the
asymmetry correction is of leading second order ${\cal
O}((\frac{g_Y}{g_N}\zeta_Y)^2)$. Thus, even in a finite nucleus where
$\zeta_Y$ may vary over the nuclear volume, we expect $R^Y_\alpha=const.$ to
a very good approximation.

In fact, these results agree with the conclusions drawn from the
analyses of single hypernuclei in the past. In the present context,
eqs.~(\ref{YNvertex}) and (\ref{RY}) are of particular interest because
they allow to extend the DDRH approach in a theoretically meaningful way
to hypernuclei using the results available already from the previous
investigations of systems without strangeness. In the applications
discussed below the nucleon (Hartree) scalar and vector vertex functions
$\Gamma_{\sigma,\omega N}(k_F)$ of \cite{FL95} will be used. The hyperon
scaling factors $R^{Y}_\alpha$ are treated as phenomenological constants to be
determined empirically. In the scalar channel information on
$R^Y_\sigma$ available from recent calculations of the J\"ulich group for
the free N$\Lambda$ T-matrix \cite{Re96,Ha98} is taken into account leaving
essentially the ratio $R^Y_\omega/R^Y_\sigma$ for $Y=\Lambda$ as the
only free parameter.

We close this section by remarking that eq.~(\ref{Gphi}) actually defines
a set of quadratic equations for the vertices $\Gamma_{\alpha B}$, as seen
immediately when inserting eq.~(\ref{DBfields}) into eq.~(\ref{Gphi}).
Vertices derived in this way would be appropriate for DDRH calculations
in Dirac-Hartree Fock approximation. From the $\sigma-\omega$ model it
is known that DHF and relativistic Hartree calculations give almost
undistinguishable results for properly adjusted parameters \cite{SW86}. In
the following we take advantage of that observation and, as in
\cite{FL95}, restrict the calculations to the Hartree case only. This
corresponds to determine the DDRH vertices by expressing the
$\Phi_\alpha$ fields on the right hand side of eq.~(\ref{Gphi}) in
Hartree approximation.

%%%%%%%%%%%%%%%%%%%%%%%%%%%%%%%%%%%%%%%%%%%%%%%%%%%%%%%%%%%%%%%%%%%%%%%%%%%%%%%
\subsection{The equations of motion}
\label{ssec:TheEquationsOfMotion}

In DDRH theory the above results are embedded into a relativistically covariant
and thermodynamically consistent field theory from which the vertices
are retrieved when evaluated in mean-field approximation.
As discussed in \cite{FL95} the $k_F$ dependence of
the DB vertices is expressed in terms of Lorentz-scalar (products of) bilinear
forms
$\hat\rho$ of the baryonic field operators $\Psi_F$. This provides a
unique mapping of the medium dependence onto frame-independent
Lorentz scalar quantities. The external
Dirac structure of the vertices is fully determined by the Lorentz character of
the meson field. The intrinsic density dependence must be deduced from
microscopic calculations as discussed in the previous section. As an obvious
generalization of the {\em ansatz} used in \cite{FL95} the DDRH vertices
are expressed here as
\beqa\label{eq:GammaIJ}
\Gamma = \Gamma_{\alpha B}(\hat{\rho}_{\alpha B}(\Psib_F, \Psi_F)),\\
\alpha = \sigma, \sigma_s, \omega, \phi, \rho  \quad
\quad B = N, \Lambda, \Sigma, \Xi \nonumber
\eeqa
where $\hat{\rho}_{\alpha B}$ denotes a Lorentz-scalar combination
of the baryon field operators.

By definition the DB vertices
$\Gamma_{\alpha B}^{DB}(k_F)$ are $c$-number valued functions
of the Hartree or Hartree-Fock expectation value of $\hat\rho_{\alpha B}^{DB}$.
From a general theoretical point of view
the DDRH vertices $\Gamma_{\alpha B}( \hat\rho_{\alpha B} )$ are not necessarily
restricted
to this particular sub-class of diagrams. Formally, a projection onto DB
correlations is
defined by the mapping \cite{FL95}
\beq\label{mapping}
\Gamma_{\alpha B}(\hat\rho_{\alpha B}) =
\int_0^\infty \Gamma_{\alpha B}^{DB}(\hat\rho_{\alpha B}^{DB})
\delta(\hat\rho_{\alpha B}^{DB} - \hat\rho_{\alpha B})
d\hat\rho_{\alpha B}^{DB} \quad .
\eeq

In the following a straightforward extension of the vector density dependence
(VDD) prescription of
ref.~\cite{FL95} will be used. This corresponds to the {\em ansatz}
\beq
\hat{\rho}_{\alpha B}[\Psib_F, \Psi_F] = \Psib_F
\hat{B}_\mu^{\alpha B}\gamma^\mu \Psi_F,
\label{eq:RhoNod}
\eeq
and chosing $\left( \hat{B}_\mu^{\alpha B} \right)_{B'B''} =
u_\mu^B \delta_{BN}\delta_{B'N}\delta_{B''N}$ with $u_\mu^B$ a four-velocity (see~\cite{FL95}).
We thus find $\hat{\rho}_{\alpha B}=\sqrt{j^B_\mu j^{B \mu}}$ and by means of
eq.~(\ref{mapping}) a Taylor series expansion of the DDRH vertex in terms
(of the modulus) of only the respective baryon four-vector current is obtained. Exactly that choice of $\hat B _\mu^{\alpha B}$ is a practical implementation of the results obtained in sect.~\ref{ssec:BBvert} This
shows that the DDRH approach in fact corresponds to expressing many-body
correlations by an expansion of vertices
into baryon n-point functions chosen such that in ground state expectation
values
correlation diagrams of the fully interacting theory are cancelled by
compensating
terms in the DDRH expansion.

As pointed out already in ref.~\cite{FL95} the most important difference
of DDRH and RMF theory are contributions from rearrangement
self-energies to the DDRH baryon field equations. Rearrangement
self-energies account physically for static polarization effects in the
nuclear medium, cancelling
certain classes of particle-hole diagrams \cite{Ne82}. Due to the
additional strangeness degree of freedom the structure of these
rearrangement self-energies is much more complex than in the purely
nucleonic DDRH theory. The variational derivative of $\Lcal_{int}$ now
leads to
\beq
\label{eq:dLIntdPsiF}
\frac{\delta\Lcal_{int}}{\delta\Psib_F} =
\frac{\p\Lcal_{int}}{\p\Psib_F} +
   \sum_{\begin{array}{c} ^{\alpha=\sigma, \sigma_s, \omega, \phi, \rho} \\
^{B=N, \Lambda,\Sigma, \Xi} \end{array}}
\frac{\p\Lcal_{int}}{\p\hat{\rho}_{\alpha B}} \frac{\delta\hat{\rho}_{\alpha
B}}{\delta\Psib_F}.
\eeq
With $S^{(r)\alpha B} \equiv \frac{\p\Lcal}{\p\hat{\rho}_{\alpha B}}$ one finds
\beqa
S^{(r)\alpha B} &=& \Psib_F \frac{\p\hat{\Gamma}_\alpha (\Psib_F,
\Psi_F)}{\p\hat{\rho}_{\alpha B}} \Psi_F \Phi^\alpha = \frac{\p\Gamma_{\alpha
B}(\hat{\rho}_{\alpha B})}{\p\hat{\rho}_{\alpha B}} \Phi^\alpha \hat{\rho}_s^B,
\quad \alpha = \sigma, \sigma_s \label{eq:Sscal}\\
S^{(r)\alpha B} &=& \Psib_F \frac{\p\hat{\Gamma}_\alpha (\Psib_F,
\Psi_F)}{\p\hat{\rho}_{\alpha B}} \gamma^\mu \Psi_F \Phi^\alpha_\mu =
\frac{\p\Gamma_{\alpha B}(\hat{\rho}_{\alpha B})}{\p\hat{\rho}_{\alpha B}}
\Phi^\alpha_\mu \hat{j}_B^\mu, \quad \alpha = \omega, \rho, \phi,\label{eq:Svec}
\eeqa
where $\hat j_\mu^B$ and $\hat\rho_s^B$ are the vector current operator and the
scalar density operator of baryon type $B$, respectively. The rearrangement
self-energy thus is given by
\beq
\hat\Sigma^{(r)} \equiv \hat\Sigma_\mu^{(r)} \gamma^\mu
\eeq
\beq
\label{eq:rearrSelfEner}
\hat\Sigma^{(r)}_\mu = - \sum_{\alpha, B} S^{(r)\alpha B} \hat B^{\alpha B}_\mu,
\eeq
where the sums in eq.~(\ref{eq:rearrSelfEner}) are those appearing in eq.~(\ref{eq:dLIntdPsiF}).
The usual self-energies \cite{SW86} are given through
\beqa
\hat\Sigma^{(0)}_s &=& \hat\Gamma_\sigma \sigma + \hat\Gamma_{\sigma_s} \sigma_s
\\
\hat\Sigma^{(0)\mu} &=& \hat\Gamma_\omega \omega^\mu + \hat\Gamma_\phi \phi^\mu
+ \hat{\vec{\Gamma}}_\rho \vec\rho^\mu + e\hat Q A^\mu,
\eeqa
where the $\hat\Gamma_\Phi$ are those defined in eq.~(\ref{eq:GammaDef}).
Thus, the total baryon self-energies are finally obtained as
\beq
\hat\Sigma_s = \hat\Sigma^{(0)}_s, \quad \hat\Sigma^\mu =
\hat\Sigma^{(0)\mu} + \hat\Sigma^{(r)\mu}.
\eeq
Here, the $\hat\Gamma_\alpha$ are diagonal matrices containing the flavour
dependent vertices.
However, in structure the baryon field equations remain unchanged
\beq
\left[\gamma_\mu \left( i\p^\mu - \hat\Sigma^\mu \right) - \left( \hat M -
\hat\Sigma_s \right) \right] \Psi_F = 0
\eeq

%%%%%%%%%%%%%%%%%%%%%%%%%%%%%%%%%%%%%%%%%%%%%%%%%%%%%%%%%%%%%%%%%%%%%%%%%%%%%%%
\subsection{Mean-field theory}
\label{ssec:MenField}

A solvable model is obtained in the Hartree mean-field approximation
which amounts to assume that products of fermion
operators are normal
ordered with respect to the Hartree ground state $\left| 0 \right>$,
given by a single slater determinant of occupied fermion states.
Expectations values with respect to the Hartree
ground state will be abbreviated as $\left< \hat O \right> \equiv
\left<0\right| \hat O \left| 0 \right>$. In the Hartree approach the
vertex functionals $\Gamma_{\alpha B}(\hat\rho_{\alpha B})$ can be treated in
a
particularly simple way. Applying Wick's theorem one gets \cite{FL95}
\beq
\left< \Gamma_{\alpha B}(\hat\rho_{\alpha B}) \right> =
\Gamma_{\alpha B}( \rho_{\alpha B} ), \qquad \rho_{\alpha B} \equiv \left<
\hat\rho_{\alpha B} \right>
\eeq
which brings the originally highly nonlinear field equations into a
tractable form. Correspondingly the rearrangement contributions are
obtained as
\beq
\left< \frac{\p\Gamma_{\alpha B}(\hat\rho_{\alpha B})}{\p\hat\rho_{\alpha B}}
\right> =
\frac{\p\Gamma_{\alpha B}(\rho_{\alpha B})}{\p\rho_{\alpha B}}.
\eeq
In the approximation as static classical fields the meson field equations reduce
to
\beqa
\left( -\nabla^2 + m_{\alpha}^2 \right) \Phi_\alpha =
\sum_{B=N,\Lambda,\Sigma,\Xi} \Gamma_{\alpha B}(\rho_{\Phi_\alpha B})
\rho^B_s, \quad  &&\alpha = \sigma, \sigma_s
\label{eq:MFScal} \\
\left( -\nabla^2 + m_{\alpha}^2 \right) \Phi_\alpha^\mu =
\sum_{B=N,\Lambda,\Sigma,\Xi} \Gamma_{\alpha B}(\rho_{\Phi_\alpha
B})j^{B\mu}, \quad  &&\alpha = \omega, \phi
\label{eq:MFVec} \\
\left( -\nabla^2 + m_\rho^2 \right) \vec\rho^\mu =
\sum_{B=N,\Lambda,\Sigma,\Xi} \vec\Gamma_{\rho B}(\rho_{\rho B})\vec j^{B\mu},
&& \label{eq:MFRho}
\eeqa
the Dirac equation for the baryons remains the only equation of motion
for an operator field, with dynamics given now by static but density
dependent self-energies
\beq
\left[ \gamma_\mu \left( i\p^\mu - \hat\Sigma^\mu (\rho) \right) - \left(
\hat M - \hat\Sigma_s \right) \right] \Psi^{MF}_F = 0.
\label{eq:MFDirac}
\eeq
In finite nuclei, where $\rho = \rho({\bf r})$, $\Sigma(\rho) =
\Sigma({\bf r})$ depends on the spatial coordinates.

%%%%%%%%%%%%%%%%%%%%%%%%%%%%%%%%%%%%%%%%%%%%%%%%%%%%%%%%%%%%%%%%%%%%%%%%%%%%%%%
%%%%%%%%%%%%%%%%%%%%%%%%%%%%%%%%%%%%%%%%%%%%%%%%%%%%%%%%%%%%%%%%%%%%%%%%%%%%%%%
\section{The $\Lambda$--$N$ model}
\label{sec:LamNMod}

In order to test the scaling relation derived in
sect.~\ref{sec:DensDepNucHyp} on existing hypernuclear data
a model with $\Lambda$-hyperons and nucleons
interacting only by non-strange mesons will be discussed. Naive quark
counting suggests that even for $\Lambda$--$\Lambda$ interactions
the strange mesons only contribute about 10\% of the interaction
strength. Keeping in mind the large uncertainties in the
hyperon--nucleon and hyperon--hyperon interactions the exchange of
strange mesons can be safely absorbed in the $\Lambda$--$\sigma$
and the $\Lambda$--$\omega$ vertices. We
will use the extended VDD prescription introduced in sect.~\ref{ssec:TheEquationsOfMotion} with a density independent
$N$--$\rho$ couplings, which leads to rather satisfactory results
for isospin nuclei \cite{FL95}.

%%%%%%%%%%%%%%%%%%%%%%%%%%%%%%%%%%%%%%%%%%%%%%%%%%%%%%%%%%%%%%%%%%%%%%%%%%%%%%%
\subsection{$\Lambda$--meson interaction}
\label{ssec:LamMesInt}

Due to the simple interaction structure of $\Lambda$ hyperons -- the
$\Lambda$s are isoscalar and electrically neutral and thus couple neither
to the $\rho$ meson nor to the Coulomb field -- the investigation of the
$\Lambda$--nucleon interaction becomes rather transparent in this model.
Assuming that the $\sigma_s$- and the $\phi$-meson are pure
$s\overline{s}$-states and therefore mainly couple to the strange quark
(due to OZI supression), they will have no significant effect in
single--$\Lambda$ hypernuclei and can thus be safely neglected.

According to sect.~\ref{ssec:BBvert} the density dependent vertices for the
$\Lambda$--N model are given by:
\beqa
\Gamma_{\sigma\Lambda} &=& R_\sigma \cdot \Gamma_{\sigma N} (\hat\rho_\Lambda)
\nonumber\\
\Gamma_{\omega\Lambda} &=& R_\omega \cdot \Gamma_{\omega N} (\hat\rho_\Lambda)
\\
\hat\rho_\Lambda &=& \hat\rho_{\sigma\Lambda} = \hat\rho_{\omega\Lambda} =
\sqrt{\hat j_\mu^\Lambda \hat j^{\Lambda\mu}},\nonumber
\eeqa
where R$_{\sigma,\omega}$ from now on denotes $R^\Lambda_{\sigma,\omega}$. The
values for
$R_{\sigma,\omega}$ will be determined in sect.~\ref{ssec:RelCoupl}.
The parameterizations of the nucleon--meson vertices
$\Gamma_{iN}(\hat\rho_N)$
are taken from \cite{HW93}. Numerically a fit with a second order polynomial in
$k_F$ to the vertices derived in \cite{BM90} from nuclear matter DBHF
self-energies is used.

%%%%%%%%%%%%%%%%%%%%%%%%%%%%%%%%%%%%%%%%%%%%%%%%%%%%%%%%%%%%%%%%%%%%%%%%%%%%%%%
\subsection{$\Lambda$ rearrangement dynamics}
\label{ssec:LamRearrDyn}

The considerations of section \ref{ssec:LamMesInt} define the dynamics of the
$\Lambda$--nucleon system, i.e. the usual and the rearrangement self-energies
can now be specified. Going to the nuclear rest frame, the arguments of the
vertex functionals, $\hat\rho_{\alpha B}$, are now defined as in
eq.~(\ref{eq:RhoNod}) with
\beq
\hat B_\mu^{B} = \hat B_\mu^{\sigma B} = \hat
B_\mu^{\omega B} = \left(
\begin{array}{cc} \hat u^N_\mu\delta^{NB} & 0 \\ 0 & \hat
u^\Lambda_\mu\delta^{\Lambda B} \end{array}
\right)
\eeq
and $\hat u_\mu = (1,0,0,0)$. This leads to the
rearrangement self-energy
\beq
\hat\Sigma_\mu^{(r)} = \sum_{B=N, \Lambda} \Psib_F \left[
\frac{\p\hat\Gamma_\sigma}
{\p\hat\rho_{\sigma B}} \sigma - \frac{\p\hat\Gamma_\omega}
{\p\hat\rho_{\omega B}} \gamma_\nu\omega^\nu \right] \Psi_F \hat B_\mu^B.
\eeq
where the nucleon and $\Lambda$ parts are given explicitly by
\beq
\Sigma_\mu^{(r)B} = \left[ \frac{\p\Gamma_{\sigma B}(\hat\rho_0^B)}{\p\hat\rho_0^B}
\sigma \hat\rho_s^B - \frac{\p\Gamma_{\omega B}(\hat\rho_0^B)}{\p\hat\rho_0^B}
\omega^\nu \hat j^B_\nu \right] u^B_\mu
\eeq
and $B = N, \Lambda$ vertices depend intrinsically only on their own densities,
as derived
in sect.~\ref{ssec:BBvert}.

%%%%%%%%%%%%%%%%%%%%%%%%%%%%%%%%%%%%%%%%%%%%%%%%%%%%%%%%%%%%%%%%%%%%%%%%%%%%%%%
\subsection{The vertex scaling factors}
\label{ssec:RelCoupl}

A consistent extension of the DDRH theory to strangeness
requires to use vertex functionals from DB
self-energies calculated in the complete octet sector. However, since such
a full scale calculation is neither
available nor feasible under the present conditions, we choose
a semi-empirical approach combining existing theoretical information on
the $\Lambda-\sigma$ vertex with a phenomenological description of the
$\omega$ vertex scaling factor.

Actually, an extension of the Bonn potential to the free $N$--$\Lambda$ system
already exists \cite{Re96,Ha98}, but DB calculations are pending. We
use the extended Bonn A potential as guideline to determine
the relative couplings R$_\sigma$ and R$_\omega$. This is consistent with
the approach used in the isospin sector because the nucleonic DDRH parameters
also were derived from the Bonn-A potential \cite{FL95}.
Because the DB interactions include highly non-linear and
non-perturbative correlation effects
the quark model reduction factor R$_q=2/3$
is not expected to be adequate for hypernuclear structure studies.

\begin{figure}
 \caption{$\chi^2$ distribution for the deviation of DDRH $\Lambda$
single particle energies and hypernuclear data (obtained in $(\pi,K)$
reactions \protect\cite{Ha96,Aj95,Pi91,Da86,Ma97}). In the calculations,
the scalar and vector vertex factors $(R_\sigma, R_\omega)$ were varied freely.}
 \label{fig:ChiSqu}
 \centering\epsfig{file=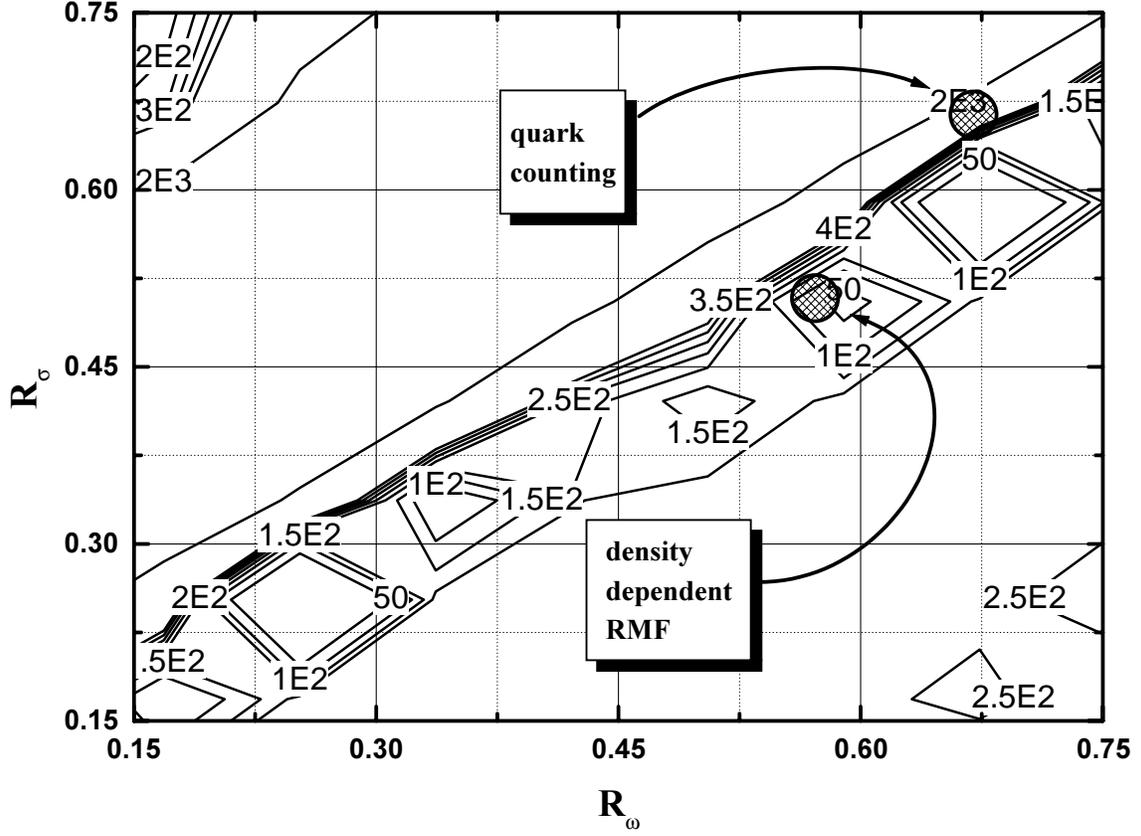,width=\linewidth}
 \end{figure}

Clearly, the final decision on the permissible $(R_\sigma, R_\omega)$
pairs is obtained from a comparision to data. In fig.~\ref{fig:ChiSqu}
the $\chi^2$ deviations of calculated and measured $\Lambda$ single
particle spectra are shown. Varying freely the R$_\sigma$ and R$_\omega$
scaling factors, DDRH single particle energies for $^{208}$Pb$_\Lambda$,
$^{89}$Y$_\Lambda$, $^{51}$V$_\Lambda$, $^{40}$Ca$_\Lambda$,
$^{28}$Si$_\Lambda$, $^{16}$O$_\Lambda$, $^{12}$C$_\Lambda$ and
$^{9}$Be$_\Lambda$ were compared to data deduced from $(\pi,K)$
\cite{Ha96,Aj95,Pi91,Da86,Ma97} experiments. With precise measurements
resolving spin-orbit doublets to high accuracy such a procedure would,
in fact, allow to fix both the scalar and vector scaling factors
unambigously because the centroid and splitting energies are determined
by difference and sum of the  scalar and vector mean-field components,
respectively.

Unfortunately, under the present experimental conditions doublets are
not resolved energetically. Typically, spin-orbit splittings are deduced
rather indirectly. e.g. by a line shape analysis \cite{Ma81} which only
allows to set constraints on the energy splitting of spin-orbit
partners. The consequences of these experimental uncertainities for a
theoretical analysis are clearly seen in fig.~\ref{fig:ChiSqu}: The
$\chi^2$ distribution is characterized by a sharp deep valley extending
between $(R_\sigma, R_\omega)=(0.1, 0.1)$ and $(R_\sigma, R_\omega) =
(0.8, 0.9)$ and any $(R_\sigma, R_\omega)$-pair in this region would
describe the data almost equally well. The quark model value pair
$R_\sigma=R_\omega=R_q$ (R$_q=\frac{2}{3}$) is seen to be located at the ridge of
valley. These results clearly illustrate the necessity of high
resolution measurements which may become possible in the near future
with a new generation of detection systems.

The probably best known case is the (1p$_{3/2}$1p$_{1/2}$) splitting in
$^{13}C_\Lambda$. A single data point is available from measurements at
Brookhaven in 1981 \cite{Ma81} from which $\Delta E = (0.36 \pm 0.3)$
MeV was deduced. In the $\chi^2$-procedure this constraint only excludes
extreme values of R$_\sigma$ and R$_\omega$ that are already ruled out
by the systematics of hyperon binding-energies (see also
sec.~\ref{ssec:SPStates}), anyway. Despite the large uncertainties the
data seem to favour a small $\Lambda$ spin-orbit splitting. Very recent
measurements at AGS/E929 \cite{Sa99} apparently confirm this conclusion.

In order to remain as close as possible to the microscopc DDRH picture
we use
\beq
R_\sigma = 0.490
\eeq
which was extracted by Haidenbauer et al. \cite{Ha98} for a sharp
$\sigma$ meson of mass $m_\sigma$=550~MeV. Because in \cite{Re96,Ha98}
the scalar meson channels were described by the correlated exchange of
pion and kaon pairs the scalar coupling also includes admixtures of a
$\sigma_s \sim s\overline{s}$ field being relevant for the $\Lambda$
couplings. Since theoretical values for the $\omega$ vertex are not
available R$_\omega$ is treated as a phenomenological parameter. From
Fig.~\ref{fig:ChiSqu} and the above value of R$_\sigma$  one finds
immediately
\beq
R_\omega = 0.553 \quad .
\eeq
In a constant coupling RMF model by Ma et al. \cite{Ma96} for
the same value of R$_\sigma$ a relative $\omega$ coupling
R$_\omega$=0.512 was obtained.  Considering the quite different DDRH interaction
structure the deviation is only apparent and, in fact, a surprisingly
good agreement can be stated. Moreover, our relative couplings are also
consistent with bounds on hyperon--nucleon couplings extracted from
neutron star models \cite{Gl92,Hu98}.

Very likely, most of the deviations of R$_\sigma$ and R$_\omega$ to the
quark model value of $2/3$ are caused by the highly nonlinear
contributions from the dynamically generated $\sigma$ and $\sigma_s$
exchange channel in \cite{Re96,Ha98}. These genuine many-body effects
superimpose additional contributions from the explicit SU(3)$_f$
symmetry breaking and $\omega-\phi$ octet-singlet mixing on the
fundamental strong interaction level.  Obviously, all these effects
cooperate in the same direction, namely to produce deviations from the
limiting values predicted by exact SU(3) symmetry. Apparently, the
attempt to represent the rather involved dynamics of the scalar channel
by a single meson of sharp mass implies an effective vector field of
compensating repulsive strength. As a consequence, neither of them can
be expected to resemble the properties of the respective bare physical
meson.

In SU(3)-symmetric models explicit symmetry breaking must be introduced
in order to reproduce hypernuclear spectra, e.g. by a symmetry breaking
term  in the Lagrangian of in the generalized chiral SU(3)
$\sigma$-model \cite{Pa99} or by means of vertex scaling factors as in
the quark-meson-coupling (QMC) model \cite{Ts98}. Because of the close
similarity of the QMC and our approach it is instructive to compare to
the results of Ref.~\cite{Ts98}: For $R_\sigma=R_q$, $R_\omega$ has to
be rescaled by a factor of 0.93, while $R_\omega=R_q$ requires to
multiply $R_\sigma$ by 1.10. As can be extracted easily from
fig.~\ref{fig:ChiSqu}, for $R_\sigma=R_q$ we would get
$R_\omega=R_q\cdot 0.97$, and, {\em vice versa}, for $R_\omega=R_q$
we find $R_\sigma=R_q \cdot 1.15$. A speciality of the QMC is a
tensor-coupling term arising from the quark-structure of the baryons
which keeps the $\Lambda$ spin-orbit splitting extremely small although
the SU(3) couplings are rather large.

%%%%%%%%%%%%%%%%%%%%%%%%%%%%%%%%%%%%%%%%%%%%%%%%%%%%%%%%%%%%%%%%%%%%%%%%%%%%%%%
%%%%%%%%%%%%%%%%%%%%%%%%%%%%%%%%%%%%%%%%%%%%%%%%%%%%%%%%%%%%%%%%%%%%%%%%%%%%%%%

\section{Relativistic Hartree description of single $\Lambda$ hypernuclei}
\label{sec:RelHarDes}

Relativistic DDRH Hartree theory and applications to isospin nuclei were
discussed in great detail in ref.~\cite{FL95} and the references
therein. Here, we present DDRH results only for single $\Lambda$
hypernuclei. The numerical realization follows closely ref.~\cite{FL95}
namely the meson fields are described by
eqs.~(\ref{eq:MFScal})--(\ref{eq:MFRho}) and baryonic wave functions are
obtained from eq.~(\ref{eq:MFDirac}). The nucleon--meson coupling
functionals are those of \cite{BM90}. The model parameters are compiled
in table~\ref{tab:ParaTab}.

%%%%%%%%%%%%%%%%%%%%%%%%%%%%%%%%%%%%%%%%%%%%%%%%%%%%%%%%%%%%%%%%%%%%%%%%%%%%%%%
\subsection{Density dependent $\Lambda$ vertices in finite nuclei}
\label{ssec:DDvert}

Numerically, DDRH calculations rely on baryon-meson vertices taken from
infinite matter DB interactions which are applied to finite nuclei in
local density approximation (LDA).
The success of DDRH theory in describing isospin nuclei
is closely related to the saturation properties of nuclear densities
from which it is clear that inifinite matter conditions are approached
gradually with increasing mass number. Under such conditions the LDA
is likely to be a rather reliable approach.
For light nuclei or, as in
single $\Lambda$ nuclei,  small fractions
of a specific baryon species with respect to the bulk components
effects from the finite size and finite particle number could
limit the applicability.

\begin{figure}
 \caption{The radial variation of the DDRH $\Lambda$--meson
vertices $\Gamma_{\sigma,\omega}$ (upper graph) and of the vector densities (lower graph) for 1s$_{1/2}$
$\Lambda$ states  in light and heavy nuclei. Results
for the single $\Lambda$ hypernuclei $^{208}$Pb$_\Lambda$ (full),
$^{16}$O$_\Lambda$ (short dashed) and $^{9}$Be$_\Lambda$ (long dashed) are
displayed.}
 \label{fig:DD-coupl}
 \centering\epsfig{file=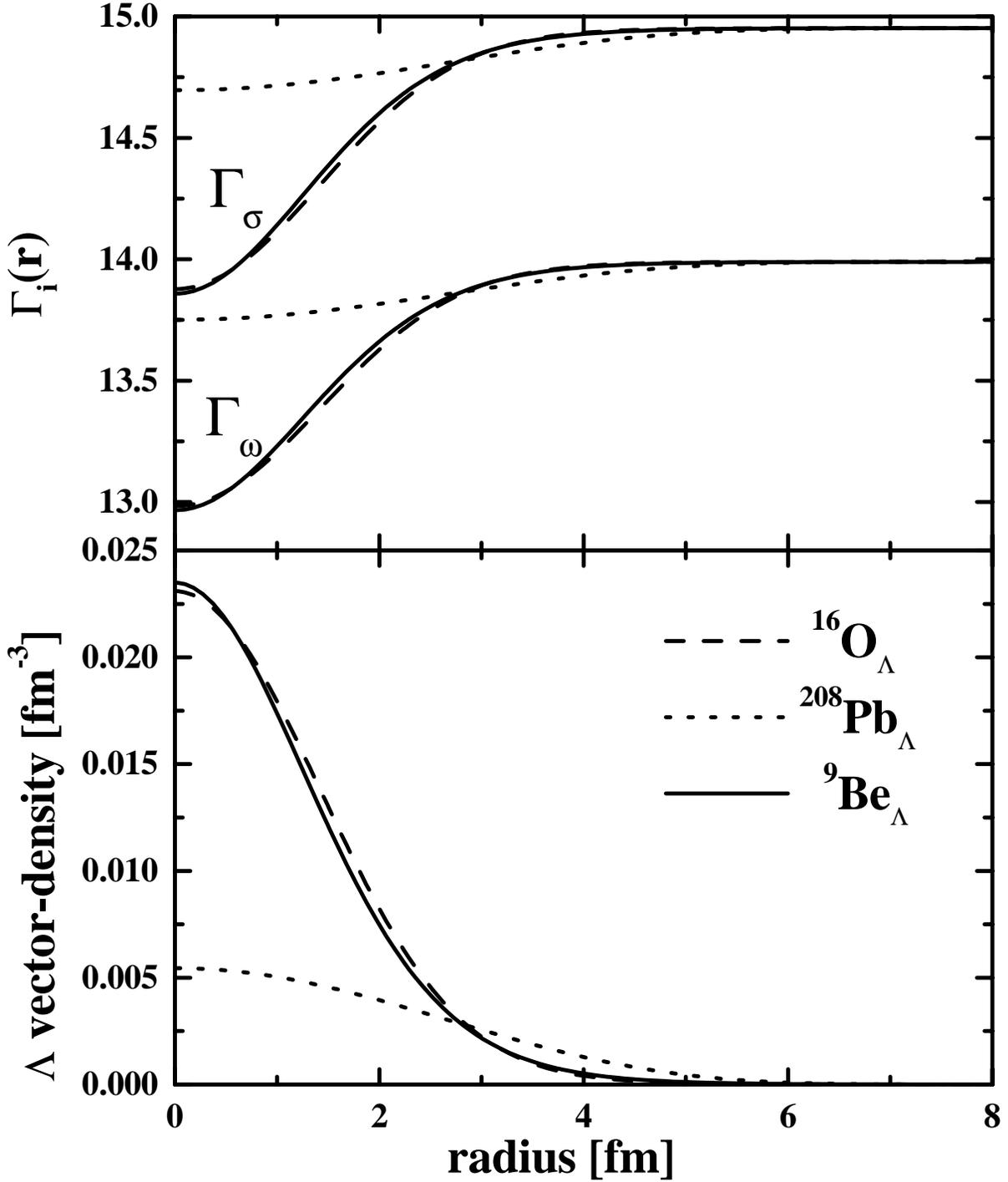,width=\linewidth}
 \end{figure}

The variation in the effective coupling strengths over the mass table is
illustrated in the upper graph of fig.~\ref{fig:DD-coupl} for the single $\Lambda$
hypernuclei $^{9}$Be$_\Lambda$, $^{16}$O$_\Lambda$ and
$^{208}$Pb$_\Lambda$. The lower graph of fig.~\ref{fig:DD-coupl} displays the vector densities for 1s$_{1/2} \Lambda$ states.
In the light nuclei the coupling decreases rapidly towards
the nuclear center while in lead the DDRH vertices are almost constant.
The behaviour follows closely the density distributions of the 1s
$\Lambda$ states. Their radial extensions are determined by the size of
the mean-field produced by the nuclear core. Of particular interest is
the variation of the vertex functionals with nuclear mass. Apparently,
the global density dependence of the infinite matter DB couplings
transforms in finite nuclei effectively into a pronounced mass number
dependence of the DDRH vertices. From these results it is clear that
a complete test of the medium dependence is only obtained in calculations
over a wide mass range. Light hypernuclei will be most appropriate to study
those
vertex parts depending explicitly on the density while heavy nuclei mainly
provide
information on hyperon interactions at a saturated density.
This also points to possible limitations of the present model: the use of
LDA vertices may lead to uncertainties in light nuclei where a particular
sensitivity on the transition to free space conditions appears.

%%%%%%%%%%%%%%%%%%%%%%%%%%%%%%%%%%%%%%%%%%%%%%%%%%%%%%%%%%%%%%%%%%%%%%%%%%%%%%%
\subsection{Structure of the $\Lambda$ mean-field}
\label{ssec:Struct}

An appropriate way to understand hypernuclear dynamics and to compare
to other calculations is to consider
the Schroedinger-equivalent potentials given in lowest order by
the difference and (the gradient of) the sum
of the relativistic scalar and vector mean-fields for the central and
the spin-orbit potentials, respectively \cite{FL95}.

\begin{figure}
 \caption{Comparison of the Schroedinger-equivalent
$\Lambda$ and neutron central potentials
including  rearrangement. Results
for $^{208}$Pb$_\Lambda$ and $^{16}$O$_\Lambda$ are shown in the upper and lower
part
of the figure, respectively. The $\Lambda$ and neutron potentials are of a
similar shape but
different depth.}
 \label{fig:nLambda-pots}
 \centering\epsfig{file=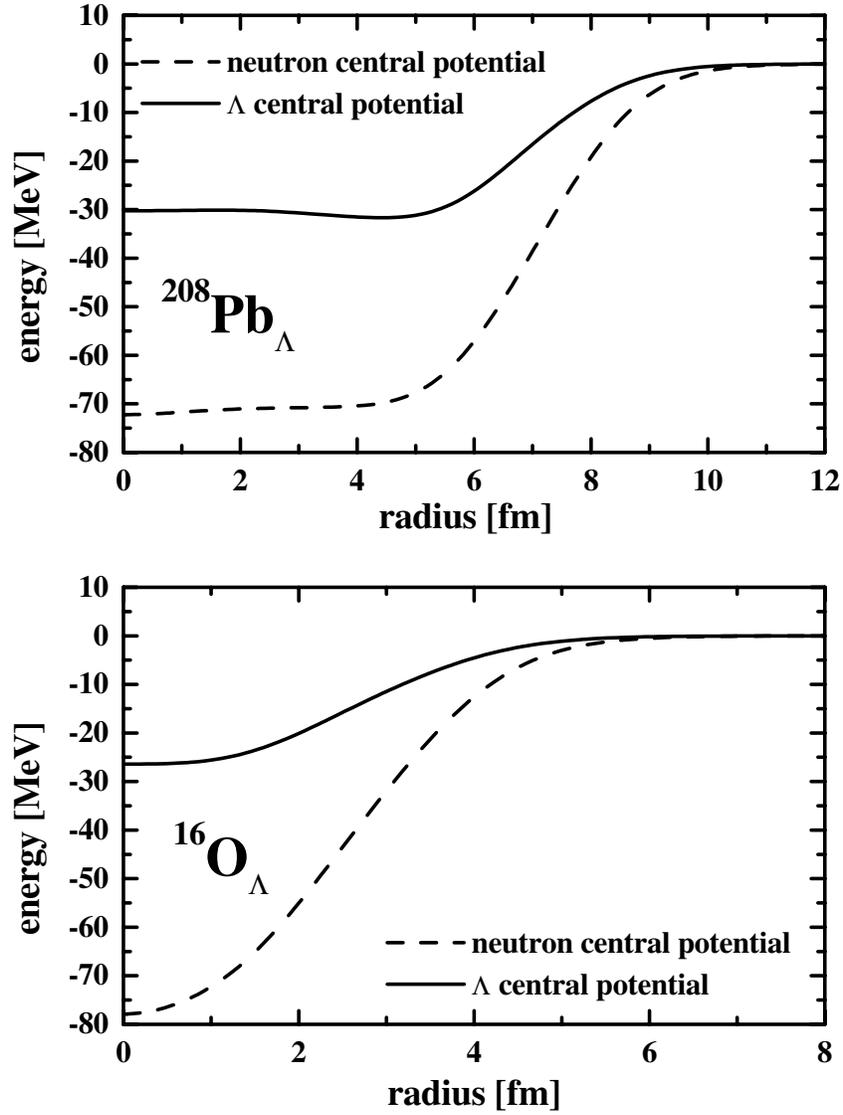,width=.7\linewidth}
 \end{figure}

\begin{figure}
 \caption{Mean-field potentials for single $\Lambda$ hypernuclei. The
lowest order Schroedinger-equivalent conventional potential (solid line)
and
rearrangement potential (dashed line) are shown. It is clearly seen that
the polarization effects described effectively by the rearrangement
potential are most important for light nuclei.}
 \label{fig:central-pots}
 \centering\epsfig{file=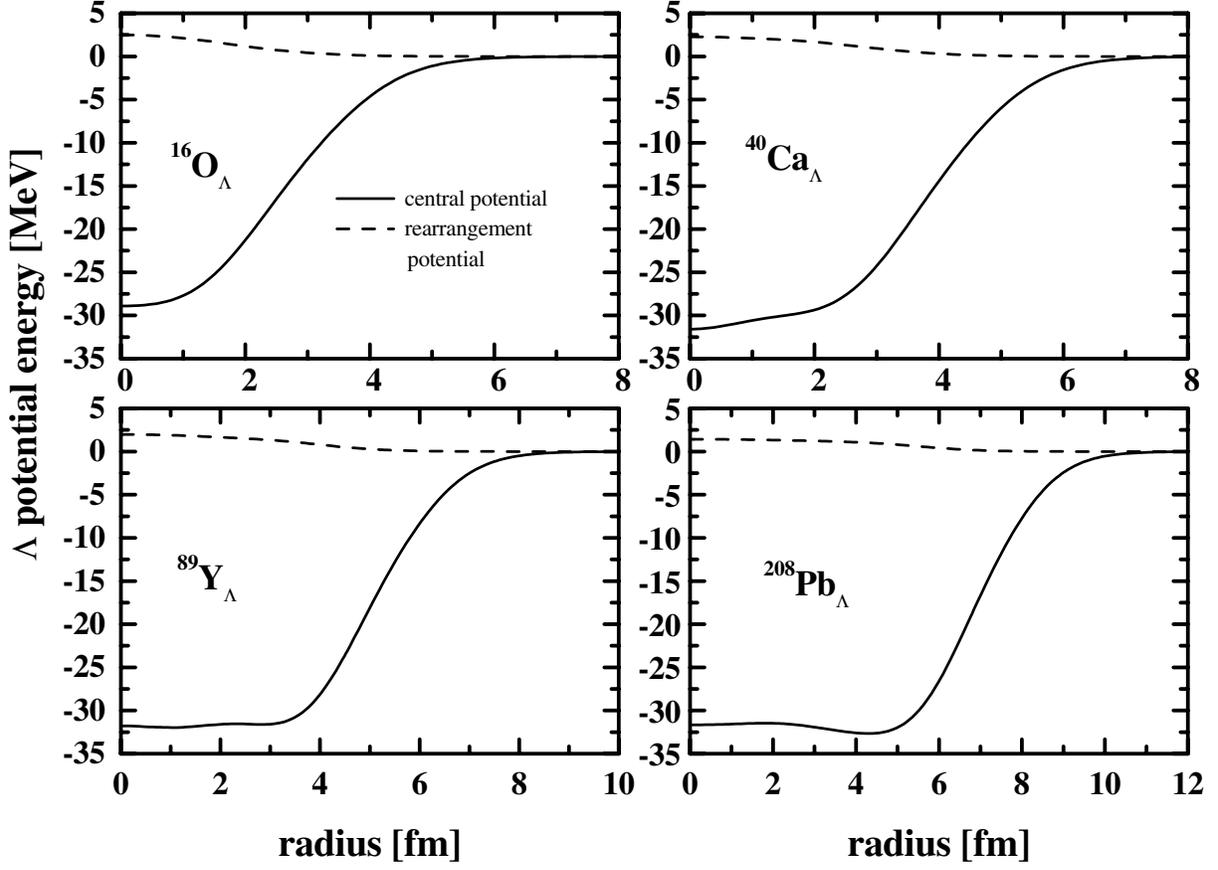,width=\linewidth}
 \end{figure}

It is obvious, that the scalar and vector fields scale according to the
$\Lambda$-meson coupling. Similar to other approaches and in agreement
with empirical anaylses the Schroedinger-type DDRH $\Lambda$-nucleus
potential is reduced by a factor of 0.35 to 0.4 compared to the nucleon
potential. Results are displayed in tab.~\ref{tab:nLambda-pots} and
fig.~\ref{fig:nLambda-pots}. Since binding energies are reduced
accordingly already the wave functions of the deepest bound $\Lambda$
states are spread over a large part of the nuclear volume resulting in a
sensitivity to the complete surrounding density structure.
Tab.~\ref{tab:rms-radii}, showing the rms-radii of $\Lambda$, neutron
and proton states in the hypernuclei $^{40}$Ca$_\Lambda$ and
$^{208}$Pb$_\Lambda$, illustrate this effect. In
fig.~\ref{fig:central-pots} the conventional central and the
rearrangement central $\Lambda$ potential for light to heavy nuclei are
displayed. It is clearly seen that the rearrangement polarization
effects are most important in the surface dominated light nuclei. Still,
the rearrangement self-energies play only a minor role for single
$\Lambda$ hypernuclei since they are weighted by the $\Lambda$ vector
density (see eqs.~(\ref{eq:Sscal}), (\ref{eq:Svec})) which is obviously
fairly small. The more important effect of the density dependent
treatment arises in hypernuclei through the nuclear core creating the
$\Lambda$'s mean-field potentials. The density dependence of the nucleon vertices
modifies the core density distribution over the whole nuclear volume
\cite{FL95}, thereby directly affecting the $\Lambda$ mean-field. Since
the potential shape and strength is  reflected in single particle
energies and wave functions a $\Lambda$ acts as an external probe
providing a global measure of the core properties.

%%%%%%%%%%%%%%%%%%%%%%%%%%%%%%%%%%%%%%%%%%%%%%%%%%%%%%%%%%%%%%%%%%%%%%%%%%%%%%%
\subsection{Single particle states}
\label{ssec:SPStates}

Hyperon single particle spectra for $|S| = 1$ hypernuclei can be seen as
a very clean fingerprint of this nucleus, since, as discussed in the
last section, they are almost undisturbed by many-body effects. Besides
the bulk structure, which contains information on the mean-field,
i.e. the nucleonic density distribution, the spectra yield
information also on other correlations of the baryonic interaction,
carried by the fine structure.

\begin{figure}
 \caption{DDRH $\Lambda$ single particle spectra in light to heavy hypernuclei.}
 \label{fig:lam-term}
 \centering\epsfig{file=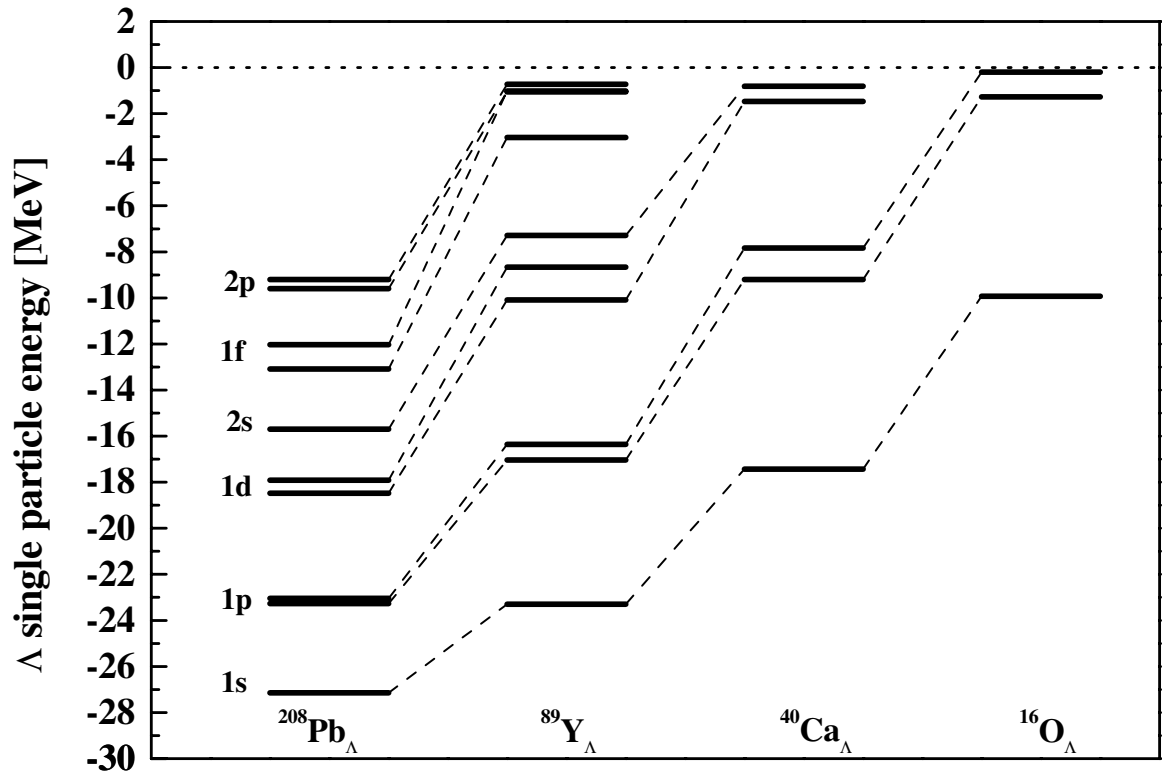,width=\linewidth}
 \end{figure}

\begin{figure}
 \caption{DDRH neutron single particle spectra in light to heavy hypernuclei.}
 \label{fig:neu-term}
 \centering\epsfig{file=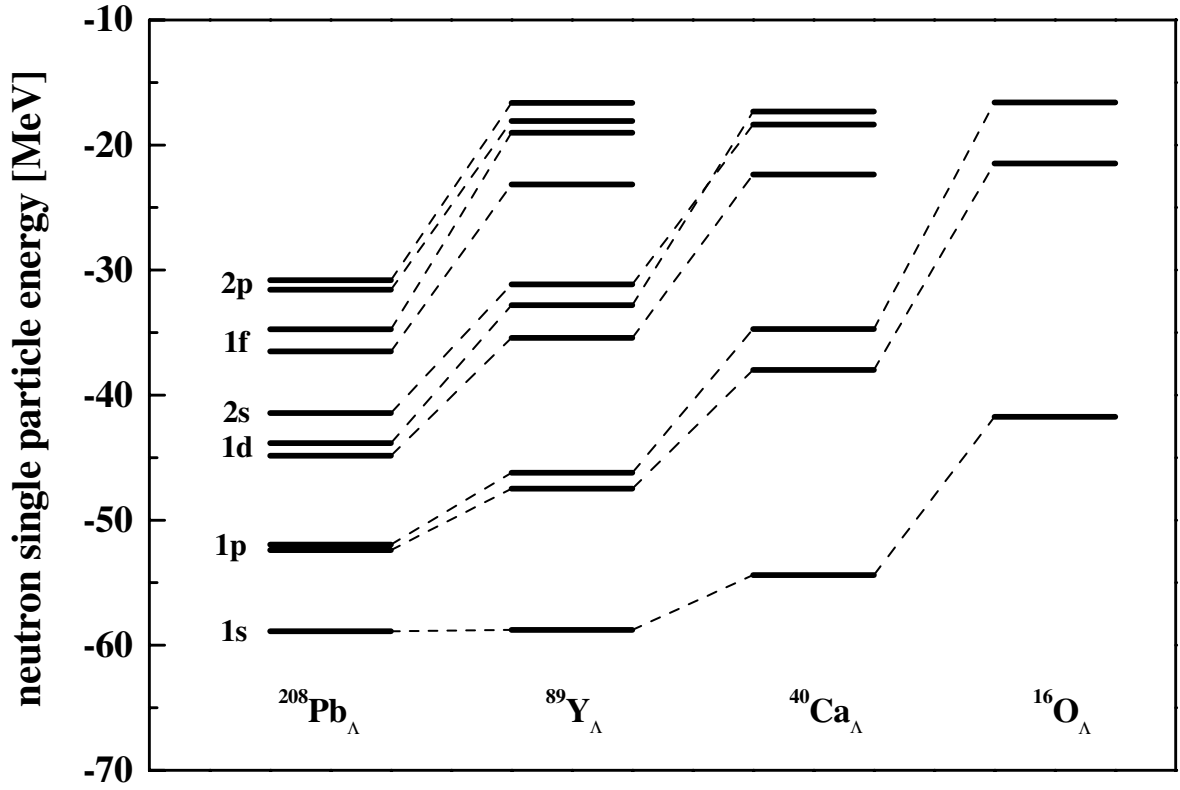,width=\linewidth}
 \end{figure}

$\Lambda$ and neutron single particle levels for light to heavy nuclei
are compared in figs.~\ref{fig:lam-term} and~\ref{fig:neu-term}. Two
major differences between the nucleonic and the $\Lambda$ spectrum are
detected:
\begin{enumerate}
\item $\Lambda$ and neutron single particle spectra are overall related
by a constant shift and an additional quenching factor because
the $\Lambda$ central potential has a depth of only about
-30 MeV, compared to -70 MeV for the neutrons (see also
fig.~\ref{fig:nLambda-pots} and tab.~\ref{tab:nLambda-pots}).
\item The
spin-orbit splitting of the $\Lambda$ states is reduced further being
less
than what is expected from the overall quenching of the potential strength.
\end{enumerate}
The DDRH calculations reproduce the experimentally suggested very small
spin-orbit splitting in $\Lambda$ hypernuclei,  e.g.~\cite{Ma81,Sa99},
rather well even without an explicit dynamical suppression of spin-orbit
interactions as e.g. a $\Lambda-\omega$ tensor coupling which is used in
the QMC model \cite{Ts98}.

\begin{figure}
 \caption{Dependence of the spin-orbit splitting for 1p$_{3/2,1/2}$, 1d$_{5/2,3/2}$,
1f$_{7/2,5/2}$, 1g$_{9/2,7/2}$ and 1h$_{11/2,9/2}$ $\Lambda$ states on
nuclear mass. The only available datapoint \protect\cite{Ma81} for the
1p$_{3/2,1/2}$ doublet in $^{13}$C$_\Lambda$ is also shown (see
sect.~\protect\ref{ssec:RelCoupl}).}
 \label{fig:so-split}
 \centering\epsfig{file=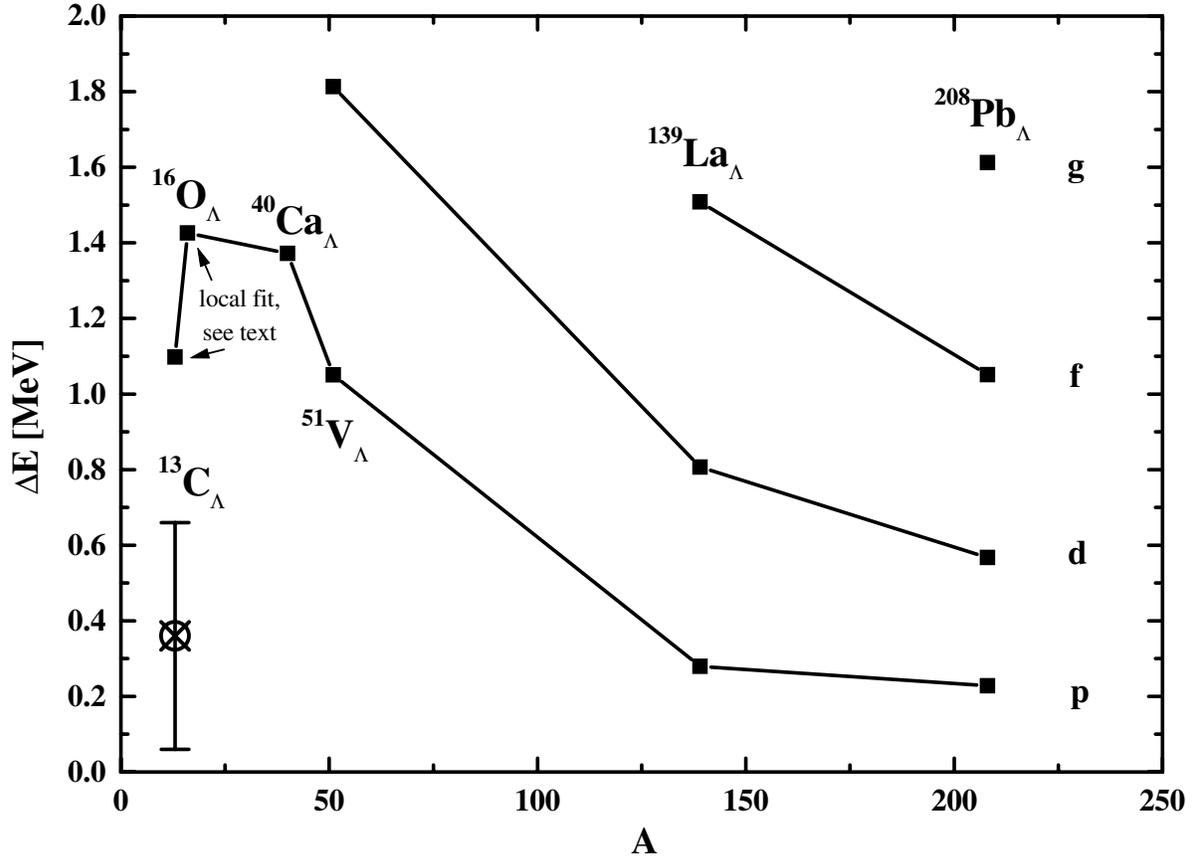,width=\linewidth}
 \end{figure}

The reason for the small spin-orbit splitting is understood by
considering the evolution with increasing mass number. In
fig.~\ref{fig:so-split} the spin-orbit splitting of $\Lambda$ states for
several nuclei across the whole mass range is shown. It can be seen
clearly that the splitting drops for higher masses what is to be
expected since it has to go to zero in the nuclear matter limit. The
splitting also drops in the low mass region -- now for the reason that
the spin-orbit doublets approach the continuum threshold and get
compressed before one of them or both become unbound. This is a
remarkable similarity to the situation found in weakly bound neutron-rich
exotic nuclei \cite{Le98}.

For exactly this reason the data point of the $^{13}C_\Lambda$ 1p shell
splitting is, though the absolute error is not that large, only of little use
to further constrain the relative coupling R$_\omega$ (see
sec.~\ref{ssec:RelCoupl}). Because the low mass hypernuclei are systematically
underbound with our standard choice for R$_{\sigma,\omega}$
the spin-orbit splittings for $^{13}C_\Lambda$ and
$^{16}O_\Lambda$ are determined with a readjusted
R$_\omega$=0.542.
\footnote{With this slightly modified R$_\omega = 0.542$
the spectra of the low mass hypernuclei are actually nicley described;
see sec.~\ref{ssec:DDvs} and
fig.~\ref{fig:spspecexpth}}.

\begin{figure}
\caption{$\Lambda$ and neutron spin-orbit potentials in $^{208}$Pb$_\Lambda$
and $^{40}$Ca$_\Lambda$. The lowest order Schroedinger-equivalent potentials are
shown.}
\label{fig:so-pot}
\centering\epsfig{file=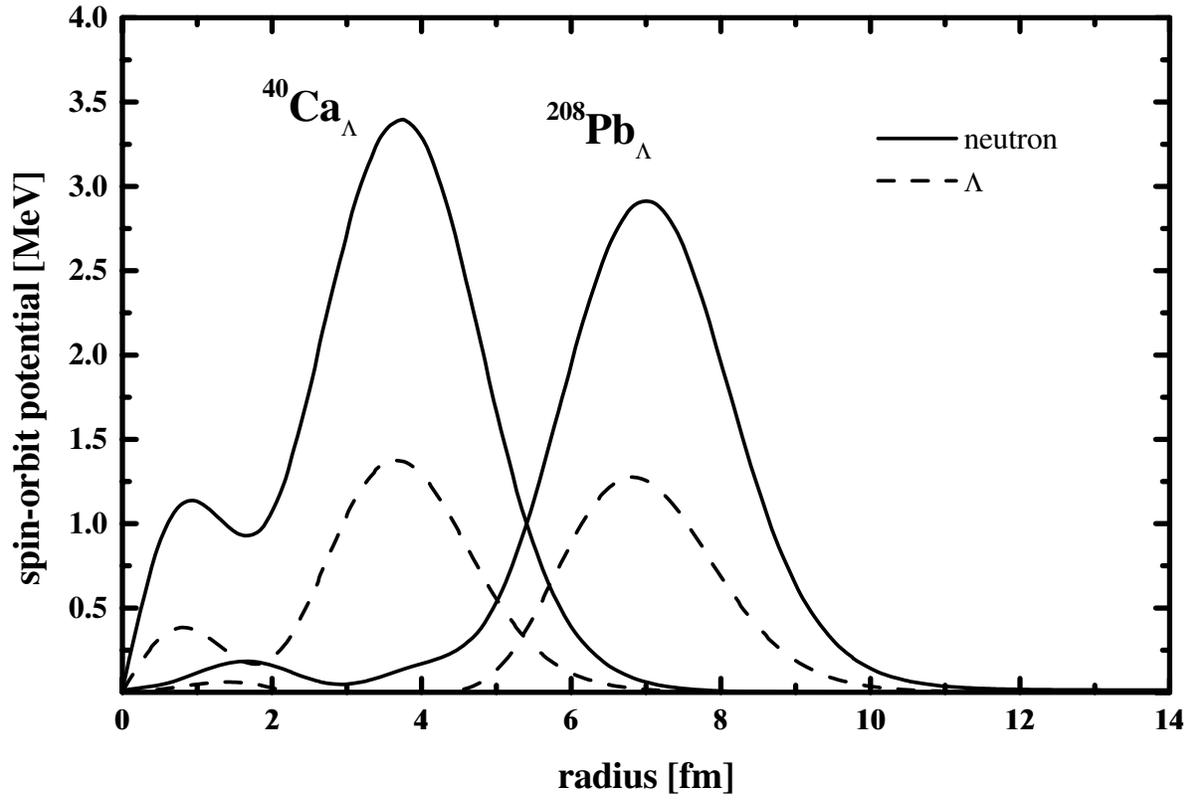,width=\linewidth}
\end{figure}

The explanation of the small spin-orbit splitting is actually found in
the overlap of the $\Lambda$ single particle wave functions and the
spin-orbit potential. As seen from binding energies and confirmed by the
r.m.s. radii of $\Lambda$ states shown in tab.~\ref{tab:rms-radii} they
are much less localized than nucleonic states in the same nucleus. The
spin-orbit potential, on the other hand, has its strongest contribution
always in a peak structure at the nuclear surface, as seen in
fig.~\ref{fig:so-pot}. Hence, the overlap of the de-localized $\Lambda$
wave functions and the rather sharply localized spin-orbit potential is
much less than for the stronger bound nucleonic wave functions. As a
result, a much smaller overlap integral and a reduced spin-orbit
interaction energy is obtained for $\Lambda$ states. The effect of the
$\Lambda$ spin-orbit potential is largest in the low angular momentum
doublets of small hypernuclei. Therefore the ideal nuclei to observe
spin-orbit splitting effects experimentally are those that are heavy
enough to bind the 1p doublet just without a too large
`close-to-threshold-squeezing'. This can be seen from
fig.~\ref{fig:so-split} to be the region around calcium.

%%%%%%%%%%%%%%%%%%%%%%%%%%%%%%%%%%%%%%%%%%%%%%%%%%%%%%%%%%%%%%%%%%%%%%%%%%%%%%%
\subsection{Systematics of single $\Lambda$ states}
\label{ssec:DDvs}

\begin{figure} \caption{Comparison of DDRH and experimental single
particle energies. The lines are drawn to guide the eye.
Results for R$_\omega$=0.542 are indicated by a dashed line.
In order to remove finite size and, especially,
surface effects, the energies are shown as a function of A$^{-2/3}$. For
A$\rightarrow$~0 they converge asymptotically to the binding energy of a
single $\Lambda$ in infinite matter, E$_\Lambda$=-28~MeV. The data
originate from $^AX (\pi, K) ^AX_\Lambda$ reactions
\protect\cite{Ha96,Aj95,Pi91,Da86,Ma97}.}
 \label{fig:spspecexpth}
 \centering\epsfig{file=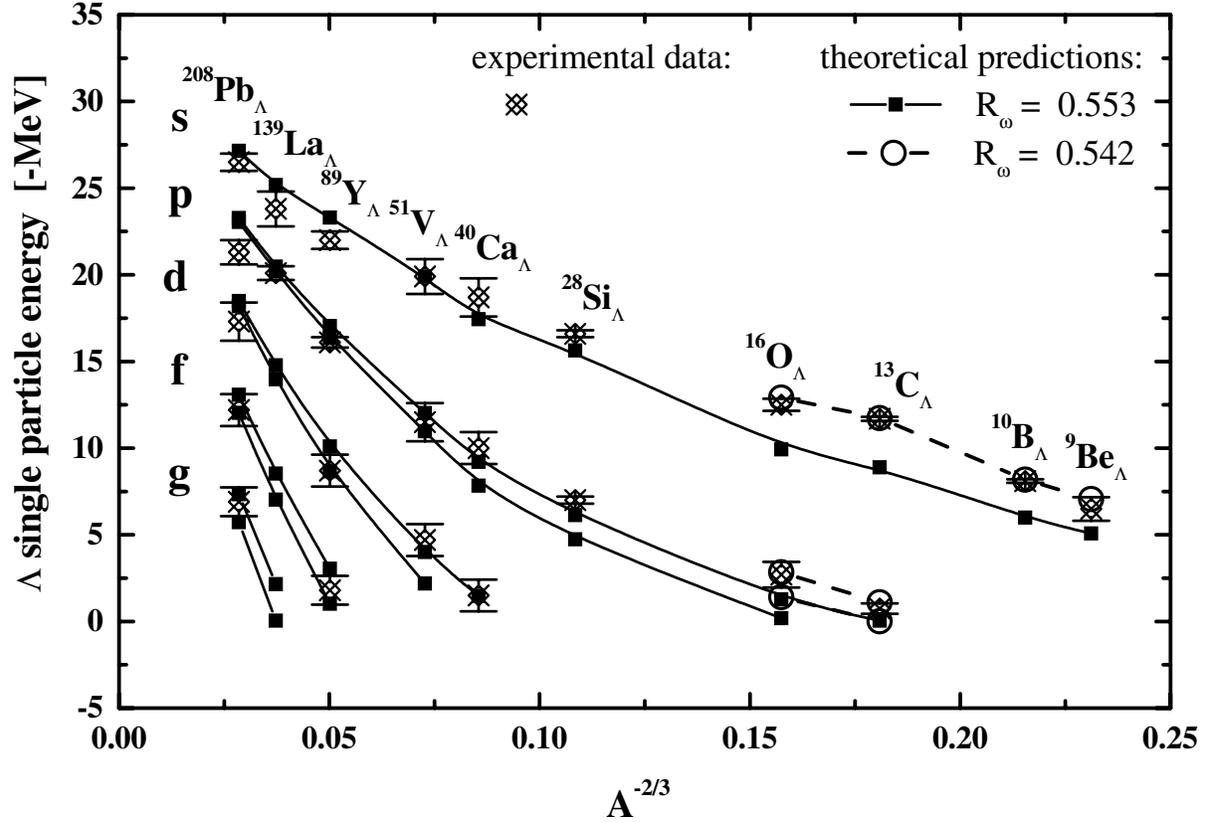,width=\linewidth}
 \end{figure}

In fig.~\ref{fig:spspecexpth} the DDRH single particle spectra are
compared to spectroscopic data from $(\pi^+, K^+)$ reactions. States in
intermediate to high mass nuclei are described fairly well by the model,
while for masses below about $^{28}$Si$_\Lambda$ deviations to the
experimental data of up to 2.5 MeV arise. As discussed in
section~\ref{ssec:DDvert}, in this mass region the limits of using LDA
vertices may be approached. We seem to overestimate systematically  the
strength of the repulsive vector interaction in low mass single
$\Lambda$ hypernuclei. This tendency already becomes apparent in going
from $^{51}$V$_\Lambda$ to $^{28}$Si$_\Lambda$. In
fig.~\ref{fig:spspecexpth} results of a calculation are included in
which  the vector scaling factor was slightly decreased by about 2\% to
R$_\omega$=0.542. Fig.~\ref{fig:ChiSqu} shows that this value is also
located in the valley of the $\chi^2$ distribution.  The $\Lambda$
separation energies in the light mass nuclei are well reproduced now but
the agreement in the high mass region, however, would deterioriate. The
result indicates the sensitivity of the DDRH calculations on fine details
of the interplay of scalar attraction and vector replusion. Clearly,
under the present conditions a 2\% variation is well within the
uncertainties of the model and, moreover, may be taken to indicate a
typical boundary for the validity of LDA in light mass nuclei.

For the heavier nuclei the microscopic DDRH results are of a quality
which is at least comparable to the phenomenological descriptions. This
we consider as a remarkable success for a model which essentially
contains only a single free parameter, namely the overall vector vertex
scaling factor R$_\omega$.

Finally, we will have a look on $(K^-,\pi^-)$ data measured at the end
of the seventies at CERN \cite{Po80,Be79}. This data set was not
included in the determination of R$_\omega$ because of its relatively
large error bars. Tab.~\ref{tab:ph-states} shows these data together
with the DDRH predictions and phenomenological RMF \cite{Ru90}
calculations. These data contain besides $\Lambda$ single particle
states also informations on the energies of $\Lambda$--particle
netron--hole states. These were calculated approximately within the
Hartree scheme by keeping a neutron hole at the specified place during
the iteration. A more realistic calculation would require to perform
a complete RPA calculation which is, however, at present neither
feasible nor worthwhile. A further complication in modelling these
nuclei are the relatively small mass numbers, as discussed already
above. Taking into acount the fairly low energy resolution of the data
and keeping in mind the previously discussed subtleties the $(K,\pi)$
data are also described satisfactorily well on a level
comparable to phenomenological RMF models.

%%%%%%%%%%%%%%%%%%%%%%%%%%%%%%%%%%%%%%%%%%%%%%%%%%%%%%%%%%%%%%%%%%%%%%%%%%%%%%%
%%%%%%%%%%%%%%%%%%%%%%%%%%%%%%%%%%%%%%%%%%%%%%%%%%%%%%%%%%%%%%%%%%%%%%%%%%%%%%%
\section{Summary and conclusion}
\label{sec:summary}

The DDRH theory introduced previously for neutron and proton isospin
nuclei was extended to hypernuclei by including the full set of
SU(3)$_f$ octett baryons. Interactions were described by a model
Lagrangian including strangeness-neutral scalar and vector meson fields
of $q\overline{q}$ (q=u,d) and $s\overline{s}$ quark character. The
medium dependence of interactions was described by meson-baryon vertices
chosen as functionals of the baryon field operators. Following DDRH
theory their structure is determined such that interaction diagrams
contributing non-perturbatively to the ground state energies and wave
functions are cancelled. Here, the DDRH vertices were chosen to cancel
Dirac-Brueckner ground state correlations. Hence, the approach
corresponds to a resummation of ladder diagrams into the vertices under
the constraint that infinite matter ground state self-energies and total
binding energies are reproduced. As the central theoretical result it
was found that the structure of Dirac-Brueckner interactions strongly
indicates that the ratio of nucleon and hyperon in-medium vertices
should be determined already by the ratio of the corresponding free
space coupling constants being affected only weakly by the background
medium. Apparently, the presently available hypernuclear data are, at
least, not contradicting such a scaling law.

Dynamical scaling will have several important consequences for
hypernuclear investigations. First of all, it  might be considered to
give a sound theoretical support to the general conviction that
hypernuclei are suitable to gain information on octett dynamics. 
A conclusion of equal importance is that hypernuclei should
follow essentially the same rules as isospin nuclei except for a shift
of energy scales. A point worthwhile to be investigated in more detail
in the future is that hypernuclear scaling might provide a way to study
the dynamical evolution of SU(3) flavour symmetry in a medium.

In order to test dynamical scaling RMF calculations for single $\Lambda$
hypernuclei were performed. Using the previously derived meson-nucleon
vertices \cite{FL95} and fixing the $\sigma\Lambda$ vertex by a theoretical
value from the literature \cite{Re96,Ha98} the $\omega$ meson-$\Lambda$ vertex
scaling factor was determined by a least square fit procedure thus
determining the only free parameter of the model empirically from a
selected set of data. Calculations over the full range of known single
$\Lambda$ nuclei led to a very satisfactory description of $\Lambda$
separation energies. The deviations from the overall agreement for
masses below A$\approx$16 are probably related to the enhancement of
surface effects in light nuclei. Very likely, they indicate the limitations of
static RMF calculations with DB vertices obtained in the local density
approximation. The minor adjustment of parameters, necessary to achieve
agreement also for the light mass systems, indicates the sensitivity of
these surface-dominated nuclei on dynamical details. In a recent
non-relativistic calculation indeed sizable contributions of hyperon
polarization self-energies especially in light nuclei \cite{Vi98} were
found.

The results are encouraging and we conclude that DDRH theory, extended
to the strangeness sector, is in fact an appropriate basis for a
microscopic treatment of hypernuclei. The present formulation and
applications are first steps on the way to a more general theory of
in-medium SU(3)$_f$ flavour dynamics. Future progress on dynamical
scaling and other theoretical aspects of the approach are depending on
the availability of Dirac-Brueckner calculations for the full baryon
octett including also the complete  pseudoscalar $0^-$ and vector $1^+$
meson multiplets.

\section*{Acknowledgement}

We thank Carsten Greiner for many useful discussions and for
pointing out open problems on the way to the final version of the
model by closely scrutinizing it.

%%%%%%%%%%%%%%%%%%%%%%%%%%%%%%%%%%%%%%%%%%%%%%%%%%%%%%%%%%%%%%%%%%%%%%%%%%%%%%%
%%%%%%%%%%%%%%%%%%%%%%%%%%%%%%%%%%%%%%%%%%%%%%%%%%%%%%%%%%%%%%%%%%%%%%%%%%%%%%%

%
% ({\it REVTEX} 3.0 automatically issues
% a \newpage command when the \begin{table} or \begin{figure}
% commands are used, so the figures and tables will be placed
% on separate pages by {\it REVTEX}).

%%%%%%%%%%%%%%%%%%%%%%%%%%%%%%%%%%%%%%%%%%%%%%%%%%%%%%%%%%%%%%%%%%%%%%%%%%%%%%
%%%%%%%%%%%%%%%%%%%%%%%%%%%%%%%%%%%%%%%%%%%%%%%%%%%%%%%%%%%%%%%%%%%%%%%%%%%%%%

\begin{table}
\caption{Model parameters of the density dependent $\Lambda$--nucleon model}
\label{tab:ParaTab}
\begin{tabular}{c|d||c|d|d}
 & & & $i$ = N & $i$ = $\Lambda$ \\ \hline \hline
 $m_N$       &      939.0 MeV & $\frac{g_{\sigma ii}^2(\rho = 0)}{4\pi}$      &
26.027 &  6.249 \\
 $m_\Lambda$ &1115.0 MeV & $\frac{g_{\sigma ii}^2(\rho = \rho_0)}{4\pi}$ &
6.781 &  1.628 \\
 $m_\sigma$  & 550.0 MeV & $\frac{g_{\omega ii}^2(\rho = 0)}{4\pi}$      &
40.240 & 12.287 \\
 $m_\omega$ & 782.6 MeV & $\frac{g_{\omega ii}^2(\rho = \rho_0)}{4\pi}$ &  9.899
&  3.022 \\
 $m_\rho$    & 770.0 MeV & $\frac{g_{\rho ii}^2}{4\pi}$                  &
1.298 &  0.000 \\
\end{tabular}
\end{table}

 \begin{table}
 \caption{Comparison of the $\Lambda$ and neutron central and central
rearrangement potential depths for single $\Lambda$ hypernuclei}
 \label{tab:nLambda-pots}
 \begin{tabular}{l|dd|dd}
    & \multicolumn{2}{c}{$^{208}Pb_\Lambda$}
    & \multicolumn{2}{c}{$^{89}Y_\Lambda$} \\
    & $\Lambda$ & $N$
    & $\Lambda$ & $N$ \\ \hline
    centr.
    & -31.7 MeV & -81.5 MeV & -31.8 MeV & -85.3 MeV \\
    c.rearr.
    & 1.4 MeV & 9.3 MeV & 2.0 MeV & 8.6 MeV \\ \hline
    $\Sigma$
    & -30.3 MeV & -72.2 MeV & -29.8 MeV & -76.7 MeV
 \end{tabular}
 \begin{tabular}{l|dd|dd}
    & \multicolumn{2}{c}{$^{40}Ca_\Lambda$}
    & \multicolumn{2}{c}{$^{16}O_\Lambda$} \\
    & $\Lambda$ & $N$
    & $\Lambda$ & $N$ \\ \hline
    centr.
    & -31.6 MeV & -90.6 MeV & -28.9 MeV & -86.5 MeV \\
    c.rearr.
    & 2.3 MeV & 10.1 MeV & 2.5 MeV & 8.6 MeV \\ \hline
    $\Sigma$
    & -29.3 MeV & -80.5 MeV & -26.4 MeV & -77.9 MeV
 \end{tabular}
 \end{table}

 \begin{table}
 \caption{r.m.s. radii of the first orbital momentum states for $\Lambda$s,
neutrons and protons in $^{40}$Ca$_\Lambda$ and $^{208}$Pb$_\Lambda$}
 \label{tab:rms-radii}
 \begin{tabular}{l|ddd|ddd}
    & \multicolumn{3}{c}{$^{40}Ca_\Lambda$}
    & \multicolumn{3}{c}{$^{208}Pb_\Lambda$} \\
    & $\Lambda$ & $n$ & $p$ & $\Lambda$ & $n$ & $p$ \\ \hline
    1s$_{1/2}$ & 2.8 fm & 2.3 fm & 2.4 fm &
    4.1 fm & 3.8 fm & 3.9 fm \\
    1p$_{3/2}$ & 3.5 fm & 3.0 fm & 3.0 fm &
    4.8 fm & 4.5 fm & 4.6 fm \\
    1p$_{1/2}$ & 3.6 fm & 3.0 fm & 3.0 fm &
    4.7 fm & 4.4 fm & 4.5 fm \\
    1d$_{5/2}$ & 4.7 fm & 3.5 fm & 3.6 fm &
    5.3 fm & 5.0 fm & 5.1 fm \\
    1d$_{3/2}$ & 6.3 fm & 3.6 fm & 3.7 fm &
    5.2 fm & 4.9 fm & 5.0 fm \\
 \end{tabular}
 \end{table}

 \begin{table}
 \caption{Transition energies for $(K,\pi)$ reactions on a nucleus
\protect\cite{Po80,Be79}. These states include $\Lambda$ particle--n hole
excitations of the single $\Lambda$ hypernuclei. The experimental values (exp.)
are compared to DDRH and a phenomenological RMF model \protect\cite{Ru90} (phen.
RMF) with nonlinear $\sigma$ self interactions.}
 \label{tab:ph-states}
 \begin{tabular}{lllddd}
    & & & exp. & DDRH & phen.RMF \\
    & n val.shell & state & [MeV] & [MeV] & [MeV] \\ \hline
    $^{12}C_\Lambda$ & $1p_{3/2}$ &
    $(1s_{1/2}\Lambda,1p_{3/2}n^{-1})$ &
    6.72$\pm$2 & 6.69 & 5.02 \\
    & & $(1p_{3/2}\Lambda,1p_{3/2}n^{-1})$ &
    18.48$\pm$2 & 15.11 & 17.21 \\ \hline
    $^{16}O_\Lambda$ & $1p_{1/2}$ &
    $(1s_{1/2}\Lambda,1p_{1/2}n^{-1})$ &
    3.35$\pm$2 & 5.76 & 3.53 \\
    & & $(1s_{1/2}\Lambda,1p_{3/2}n^{-1})$ &
    9.90$\pm$2 & 10.13 & 9.46 \\
    & & $(1p_{1/2}\Lambda,1p_{1/2}n^{-1})$ &
    13.20$\pm$2 & 16.16 & 13.89 \\
    & & $(1p_{3/2}\Lambda,1p_{3/2}n^{-1})$ &
    19.20$\pm$2 & 18.40 & 18.88 \\ \hline
    $^{40}Ca_\Lambda$ & $1d_{3/2}$ &
    $(1p_{1/2}\Lambda,1d_{3/2}n^{-1})$ &
    5.79$\pm$2 & 8.84 & 7.40 \\
    & & $(1d_{3/2}\Lambda,1d_{3/2}n^{-1})$ &
    14.47$\pm$2 & 11.34 & 15.48 \\
    & & $(1d_{5/2}\Lambda,1d_{5/2}n^{-1})$ &
    19.35$\pm$2 & 20.07 & 20.71
 \end{tabular}
 \end{table}

%%%%%%%%%%%%%%%%%%%%%%%%%%%%%%%%%%%%%%%%%%%%%%%%%%%%%%%%%%%%%%%%%%%%%%%%%%%%%%%
%%%%%%%%%%%%%%%%%%%%%%%%%%%%%%%%%%%%%%%%%%%%%%%%%%%%%%%%%%%%%%%%%%%%%%%%%%%%%%%

 \pagebreak

 \pagebreak

 \pagebreak

 \pagebreak

 \pagebreak

 \pagebreak

 \pagebreak

\pagebreak

\end{document}